\documentclass[a4paper,12pt]{article}
\usepackage{latexsym,amsmath,amsfonts,amssymb}
\usepackage{multirow}
\usepackage[dvips]{graphicx}
\usepackage{mathrsfs,mathtools}
\usepackage{pstricks,pst-node,pst-text,pst-3d}
\usepackage{slashed}
\usepackage{verbatim}
\usepackage{feynmp}
\usepackage[nosort]{cite}
\usepackage{hyperref}
\definecolor{mypink1}{rgb}{0.858, 0.188, 0.478}
\definecolor{xanadu}{rgb}{0.45, 0.53, 0.47}
\definecolor{amethyst}{rgb}{0.6, 0.4, 0.8}
\definecolor{green}{rgb}{0.29, 0.33, 0.13}
\definecolor{brightlavender}{rgb}{0.75, 0.58, 0.89}
\definecolor{brown(traditional)}{rgb}{0.59, 0.29, 0.0}
\definecolor{cadetblue}{rgb}{0.37, 0.62, 0.63}
\definecolor{capri}{rgb}{0.0, 0.75, 1.0}
\definecolor{coolgrey}{rgb}{0.55, 0.57, 0.67}
\definecolor{darkcyan}{rgb}{0.0, 0.55, 0.55}
\definecolor{darkgreen}{rgb}{0.0, 0.2, 0.13}
\definecolor{darkmagenta}{rgb}{0.55, 0.0, 0.55}
\definecolor{darktan}{rgb}{0.57, 0.51, 0.32}
\definecolor{darkviolet}{rgb}{0.58, 0.0, 0.83}
\definecolor{fuzzywuzzy}{rgb}{0.8, 0.4, 0.4}
\definecolor{frenchlilac}{rgb}{0.53, 0.38, 0.56}
\definecolor{lilac}{rgb}{0.78, 0.64, 0.78}
\definecolor{phlox}{rgb}{0.87, 0.0, 1.0}
\definecolor{darkraspberry}{rgb}{0.53, 0.15, 0.34}
\definecolor{darkred}{rgb}{0.55, 0.0, 0.0}
\definecolor{deepcerise}{rgb}{0.85, 0.2, 0.53}

\newcommand{\tr}[1]{\text{Tr}\big[#1\big]}

\newcommand{\sect}[1]{\setcounter{equation}{0}\section{#1}}

\begin{document}

\begin{titlepage}

\setcounter{page}{0}

\begin{flushright}
{ }
\end{flushright}

\vspace{0.6cm}

\begin{center}
  {\Large \bf A new higher-spin theory of supergravity \\ in $2+1$ dimensions}

\vskip 0.99cm

{\bf
George Georgiou$^a$
}
\vskip 1.5cm
{\em
${}^a$Demokritos National Research Center, Institute of Nuclear and Particle Physics,
Ag. Paraskevi, GR-15310 Athens, Greece,\\
}

\vskip 0.8cm
{\small \sffamily
georgiou@inp.demokritos.gr
}

\vskip 1.5cm

\end{center}

\begin{abstract}
We construct a novel higher-spin theory of gravity in $2+1$ spacetime dimensions. The construction is based on a higher-spin super-algebra extending the Poincare  group.
Our algebra accommodates all integer and half-integer spins from $1$ to infinity and, in contrast to the results in the existing literature,
allows for an infinite number of particles of spin $s$.
Subsequently, we generalise the construction to include a cosmological constant. In this case, the higher-spin group is an extension of the AdS or dS group and contains only bosonic generators.
Treating the higher-spin groups as  gauge groups we write down the Chern-Simons like action,
the transformation laws for the fields of the theory and their equations of motion  in each of the three aforementioned cases.
Finally, we comment on the generalisation of our algebras in $d+1$ dimensions.

\end{abstract}

\vfill

\end{titlepage}

\sect{Introduction}\label{sec:intro}
All the fields and their corresponding particles, observed in nature so far, have spin less or equal to two. 
Similarly, our best theoretical description of nature is in terms of field theories based on the gauge principle 
and predict particles with spin  less or equal to two, as well. However, it would be desirable, both aesthetically and 
mathematically to construct theories which can accommodate particles with spin higher than two.
On one hand, our best candidate for a consistent quantisation of gravity, superstring theory predicts towers of particles 
whose spin is unbounded. Despite this fact, all these particles of higher spin acquire large masses and essentially decouple from 
the observed particles at the low energy regime \footnote{In general, large spin operators play a prominent role both at the deep inelastic scattering in QCD, as well as in the context of the AdS/CFT correspondence  \cite{Georgiou:2010zt}.}.
On the other hand, there is no a-priory reason why one can not write down more general actions which describe massless higher-spin particles. This has been achieved in the free theory limit (no interactions) in the work of Fronsdal  \cite{Fronsdal:1978rb}. The actions one can write 
down are invariant under generalised gauge transformations of the higher spin fields.  Despite this success, one encounters dire difficulties when one tries to include interactions of the higher-spin fields with the particles of spin equal or less than two \cite{Deser}. Consistent cubic vertices of higher-spin fields interacting among themselves around flat space-time were obtained for the first time in \cite{Bengst}.  \\The problem of introducing consistent interactions 
between higher-spin fields and gravity in four dimensions was partially solved  in \cite{Fradkin:1987ks}. 
To avoid several no-go theorems \cite{nogo} the authors relaxed the condition that the theory is formulated around flat spacetime -they considered expansion around $AdS$ spacetimes- in which case the  $S$-matrix is not well-defined.
Furthermore, they dropped the assumption that the spectrum contains only a finite number of particles. However, there was an obstacle in going beyond the cubic level basically because it is difficult to generalise certain 
constraint equations needed to solve for the so-called 'extra fields' in terms of the dynamical fields of the theory.  

In the recent years, a huge effort has been made in the direction of studying interactions of higher-spin ($s>2$) particles among themselves and with lower spin ($s\leq 2$) fields. Fully non-linear equations of motion and cubic vertices
for the higher-spin fields were obtained
in \cite{Vasiliev:1990en}.  
The progress made is nicely summarised in the reviews \cite{rev}.
Furthermore, the interactions of non-abelian tensor gauge fields with themselves and with the gluon were also studied-see \cite{Savvidy:2010kw} and references therein.
Finally, various aspects regarding the precise connection of higher-spin theories with string theory and holography
have been analysed in \cite{stringy,ads/cft}. 

In this work we will focus on higher-spin interacting theories in $2+1$ spacetime dimensions.
Our starting point will be the result of references \cite{Townsend,Witten} which states that $2+1$ dimensional gravity  is a gauge theory described by a Chern-Simons action with a connection one-form
taking values in the Lie algebra of isometries of the vacuum solution which will chosen to be either Minkowski, $AdS$ or $dS$. The relevant Lie algebras  will correspondingly be  $ISO(2,1)$, $SO(2,2)$ or $SO(3,1)$.
For some comprehensive reviews on the subject see \cite{Kiran}.

The plan of the paper is as follows. In Section 2 we consider the novel higher-spin superalgebra which extends the $ISO(2,1)$ group.
After providing the reader with the (anti)commutation relations defining the superalgebra, we give the invariant quadratic form which 
this superalgebra admits. We then proceed to show that the generators of the superalgebra satisfy the necessary graded Jacobi identities.
In Section 3, equipped with this superalgebra, we write down a Chern-Simons like action that is invariant under the aforementioned higher-spin group, 
the equations of motion for the fields of the theory and their transformation laws. We also comment on the different structure of our theory compared to 
the higher-spin theories existing in the literature.
In Section 3, we generalise our construction to include a negative ($AdS_3$) or positive ($dS_3$) cosmological constant. 
Finally, in Section 4, we conclude and discuss possible directions for future research.
In the first of the two Appendices we prove that our algebras admit the specific invariant forms that are given in the main text,
while in the second Appendix we assemble our notations for supersymmetry in $2+1$ dimensions.

\bibliographystyle{nb}

\sect{A new higher-spin algebra }\label{sec:algebra}
As mentioned in the Introduction, it is highly non-trivial to write down gauge invariant actions involving higher spin particles interacting minimally with gravity.
The main reason is that, although this is possible at the free level, upon covariantisation the derivatives do not commute resulting to a non-invariant action with respect
to the higher-spin gauge symmetries \cite{rev}.

One of the basic assumptions when constructing the algebra is that there exists only one particle per spin $s$ (hypothesis II of \cite{Fral}).
In this Section we show that by relaxing this condition it is possible to write down a new higher-spin algebra that can serve as the basis for constructing higher-spin invariant
actions coupled to gravity. We start by giving the result. We then proceed to prove that the given algebra satisfies the Jacobi identities and admits a certain invariant bilinear form.\\
To begin with, we consider the gauge field in $2+1$ spacetime dimensions. This can be written as
\begin{eqnarray}\label{gauge}
\Omega_\mu= \sum_{n=1}^{\infty} \frac{1}{n!}e_\mu^{a_1a_2...a_n}P_{a_1a_2...a_n} +\sum_{n=1}^{\infty} \frac{1}{n!} \omega_\mu^{a_1a_2...a_n}J_{a_1a_2...a_n}.
\end{eqnarray}
In \eqref{gauge} we have introduced an infinite number of gauge fields which are fully symmetric but not traceless with respect to the tangent space indices $a_1,a_2,...,a_n$.
The same symmetry property is shared by the generators $P_{a_1a_2...a_n}$ and $J_{a_1a_2...a_n}$. One can recognise the vielbein $e_\mu^{a}$ and spin connection $\omega_\mu^{a}$ as the first terms in the expansion of $\Omega_\mu$ \footnote{Notice that in 2+1 dimensions the spin-connection can be written with a single tangent space index since one can replace $J^{ab}$ with $J^a=1/2 \epsilon^{abc} J_{bc}$.}. The subsequent terms in the expansion will be the fields describing the higher-spin "particles".

We now proceed to write down the algebra satisfied by the bosonic generators $P_{a_1a_2...a_n}$
and $J_{a_1a_2...a_n}$. The commutation relations for the generators read
\begin{eqnarray}\label{JJcom}
[J_{a_1a_2...a_{n_1}}, J_{b_1b_2...b_{n_2}}]=
\sum_{i=1}^{n_1}\sum_{j=1}^{n_2}\epsilon_{a_i b_j c}J^c_{a_1...\hat{a}_i...a_{n_1} b_1...
\hat{b}_j ...b_{n_2}}=\epsilon_{a_1b_1c}J^c_{a_2...a_{n_1} b_2...b_{n_2}}+...
\end{eqnarray}
\begin{eqnarray}\label{JPcom}
[J_{a_1a_2...a_{n_1}}, P_{b_1b_2...b_{n_2}}]=
\sum_{i=1}^{n_1}\sum_{j=1}^{n_2}\epsilon_{a_i b_j c}P^c_{a_1...\hat{a}_i...a_{n_1} b_1...
\hat{b}_j ...b_{n_2}}=\epsilon_{a_1b_1c}J^c_{a_2...a_{n_1} b_2...b_{n_2}}+...\
\end{eqnarray}
\begin{eqnarray}\label{PPcom}
[P_{a_1a_2...a_{n_1}}, P_{b_1b_2...b_{n_2}}]=0,
\end{eqnarray}
where the hat appearing in the right hand side of \eqref{JJcom} and \eqref{JPcom} is to denote that
the corresponding index is missing from the generator and the dots denote all possible permutations
for each set of indices ${a_1,a_2,...,a_{n_1}}$ and ${b_1,b_2,...,b_{n_2}}$.
We would also like to point out that the generators  in the right hand side of \eqref{JJcom} and \eqref{JPcom}  are fully symmetric with respect to all their indices once the index $c$ is brought down
with the tangent space metric $\eta_{ac}=diag(-1,1,1)$. Finally, let us note that the algebra written above has as a subalgebra the Poincare group in $2+1$ dimensions $ISO(2,1)$ which is generated by
$J_a$ and $P_a$. We will be calling this extended algebra $hISO(2,1)$.

A couple of important comments are in order. The first thing to notice is that the spectrum contains infinite towers of higher-spin fields which are not legitimate to truncate at some finite spin $s$.
This can be seen from the structure of the algebra \eqref{JJcom},  \eqref{JPcom} and \eqref{PPcom}.
Secondly, due to the fact the generators of the algebra are not traceless
there is an infinite set of generators that correspond to particles with spin-s. One could, in principle,
decompose each of the generators  $J_{a_1a_2...a_{s}}$ to its irreducible representations under the
Lorentz group but we will not attempt this since the algebra in terms of the generators transforming
under irreducible representations is tedious.
We should stress that
one should treat the generators $J_{a_1a_2...a_{s}}$, $J^b_{b a_1a_2...a_{s}}$,
$J^{b_1b_2}_{b_1 b_2a_1a_2...a_{s}}$ and their corresponding fields $e_{\mu}^{a_1a_2...a_{s}}$, $e_{\mu\, b}^{\,\,\,\,\,b a_1a_2...a_{s}}$ , $e_{\mu\, b_1b_2}^{\,\,\,\,\,b_1b_2a_1a_2...a_{s}}$, and so on, as different objects each of which is describing a multiplet
of particles with spins from $1$ to $s+1$, for $s$ even or with spins from $2$ to $s+1$, for $s$ odd. We should mention that such a proliferation of particles of spin $s$ has also been observed
in the \cite{Antoniadis:2011re}.

At this point we should stress the fundamental difference between our algebra and those that exist
in the literature \cite{Ble,Fral}. In the latter each generator with $s$ indices is traceless and unique and as a result
it appears in the right hand side of an infinite number of commutation relations. This happens because
the anti-commutator of two irreducible representations is no longer irreducible. As a result the traceless generator with $s$ indices appears in the right hand side of anti-commutators whose total number of indices is $s+2,s+4,s+6,...$ (see equation (5.12) of the first reference of \cite{rev}).
In contradistinction, in our algebra any generator appears only in a finite number of anti-commutation relations (see \eqref{JJcom} and \eqref{JPcom}).

Furthermore, as shown in the Appendix this higher spin algebra admits an invariant scalar product of the  form
\begin{eqnarray}\label{flatform}
&&\langle J_{a_1a_2...a_{k}},\,\,\,P_{a_{k+1}a_{k+2}...a_{2s}}\rangle=\sum_{p}\eta_{p(a_1)p(a_2)}\eta_{p(a_3)p(a_4)}...\eta_{p(a_{2s-1})p(a_{2s})} \nonumber \\
&& \langle J_{a_1a_2...a_{k}},\,\,\,P_{a_{k+1}a_{k+2}...a_{2s+1}} \rangle=0, \langle J_{a_1a_2...a_{n_1}},\,\,\,J_{b_{1}b_{2}...b_{n_2}} \rangle=0 \nonumber \\
&&\langle P_{a_1a_2...a_{n_1}},\,\,\,P_{b_{1}b_{2}...b_{n_2}} \rangle=0
\end{eqnarray}
where the sum in the first equation of \eqref{flatform} is over all possible inequivalent permutations of the indices
$a_1,a_2,...,a_{2s}$. Two permutations are regarded as equivalent when one can be obtained from the other by exchanging the indices which belong to the same etas, e.g. $\eta_{a_1a_2}\eta_{a_3a_4}...\sim \eta_{a_2a_1} \eta_{a_3a_4}...\sim \eta_{a_1a_2} \eta_{a_4a_3}... \sim \eta_{a_2a_1} \eta_{a_4a_3}...  $.

\subsection{A supersymmetric extension}
In this Section, we show that it is possible to generalise the bosonic higher spin algebra of the previous Section to an algebra with a $Z_2$ grading. This new algebra includes an infinite set of fermionic generators $Q_{\alpha\,a_1a_2...a_s}$ and has as a subalgebra the usual supersymmetric algebra
in $2+1$ dimensions \eqref{susy3d}. 
We will be calling this extended superalgebra $shISO(2,1)$.
 We should mention that the index $\alpha$ carried by these fermionic generators is a spinor index. In general we reserve the Greek alphabet to denote the spinor indices while we use the Latin alphabet to denote tangent space indices. For the curved space-time indices
we will be using the letters $\mu, \nu,\kappa, \lambda, \rho, \sigma$.

Obviously, the bosonic algebra of  \eqref{JJcom}, \eqref{JPcom} and \eqref{PPcom}, namely  $hISO(2,1)$ is also a subalgebra of this extended by fermionic generators algebra.
The fermionic generators
correspond to half-integer particles and together with the bosonic ones can be assembled in
a super-connection field which reads
\begin{eqnarray}\label{gaugesusy}
\Omega_\mu= \sum_{n=1}^{\infty} \frac{1}{n!}e_\mu^{a_1a_2...a_n}P_{a_1a_2...a_n} +\sum_{n=1}^{\infty} \frac{1}{n!} \omega_\mu^{a_1a_2...a_n}J_{a_1a_2...a_n}+\sum_{n=0}^{\infty} \frac{1}{n!}
\psi_\mu^{\alpha\, a_1a_2...a_n}Q_{\alpha\, a_1a_2...a_n} .
\end{eqnarray}
After introducing the necessary fields we are in position to write down the (anti-)commutation relations
for this novel higher spin superalgebra. The anticommutators of the bosonic  subalgebra is given by
\eqref{JJcom},  \eqref{JPcom} and \eqref{PPcom}. The (anti-)commutation relations involving the infinite tower of fermionic generators read
\begin{eqnarray}\label{QQcom}
\{Q_{\alpha\,a_1a_2...a_{n_1}},Q_{\beta\,b_1b_2...b_{n_2}} \}=
2 i \,(\gamma^c)_{\alpha\beta}P_{c a_1a_2...a_{n_1} b_1b_2...b_{n_2}}-2i \sum_{i=1}^{n_1}(\gamma_{a_i}) _{ \alpha\beta}P_{a_1...\hat{a}_i...a_{n_1}b_1b_2...b_{n_2}}\nonumber \\
-2i \sum_{i=1}^{n_2}(\gamma_{b_i}) _{ \alpha\beta}P_{a_1a_2...a_{n_1}b_1...\hat{b}_i...b_{n_2}}-4 i\sum_{i=1}^{n_1}\sum_{j=1}^{n_2}\epsilon_{\alpha\beta} \epsilon_{a_i b_j c} P^c_{a_1...\hat{a}_i...a_{n_1} b_1...\hat{b}_j...b_{n_2}}.
\end{eqnarray}
In the case where one or both of the $n_1$ and $n_2$ are zero the last term of \eqref{QQcom} will not be present. If $n_1=0$ the second term in the right hand side of the aforementioned equation will be missing too, while if $n_2=0$ the third term will be absent.\\
The anti-commutation relations of the supercharges with the bosonic generators take the form
 \begin{eqnarray}\label{QJcom}
[J_{a_1a_2...a_{n_1}},Q_{\alpha\,b_1b_2...b_{n_2}}]=\sum_{i=1}^{n_1}\sum_{j=1}^{n_2}\epsilon_{a_i b_j c}Q^{\,\,\,\,\,c}_{\alpha \,a_1...\hat{a}_i...a_{n_1} b_1...
\hat{b}_j ...b_{n_2}}-\frac{1}{2} \sum_{i=1}^{n_1}(\gamma_{a_i})_{\alpha}^ {\,\,\,\,\beta} Q_{\beta \,a_1...\hat{a}_i...a_{n_1}b_1b_2...b_{n_2}}
\end{eqnarray}
\begin{eqnarray}\label{QPcom}
[P_{a_1a_2...a_{n_1}},Q_{\alpha\,b_1b_2...b_{n_2}}]=0.
\end{eqnarray}
As in the bosonic subalgebra, the higher-spin superalgebra $hsISO(2,1)$ defined by equations \eqref{QQcom}, \eqref{QJcom}, \eqref{QPcom}, \eqref{JJcom},  \eqref{JPcom} and \eqref{PPcom} does not admit a consistent truncation up to some finite spin $s$. 

Finally, let us write down the invariant scalar product for the case of the $Z_2$-graded algebra
\begin{eqnarray}\label{flatformsusy}
&&\langle J_{a_1a_2...a_{k}},\,\,\,P_{a_{k+1}a_{k+2}...a_{2s}}\rangle=\sum_{p}\eta_{p(a_1)p(a_2)}\eta_{p(a_3)p(a_4)}...\eta_{p(a_{2s-1})p(a_{2s})} \nonumber \\
&&\langle J_{a_1a_2...a_{k}},\,\,\,P_{a_{k+1}a_{k+2}...a_{2s+1}} \rangle=0,\langle J_{a_1a_2...a_{k}},\,\,\,Q_{\alpha \, a_{k+1}a_{k+2}...a_{2s+1}}\rangle=0 \nonumber \\
&&\langle J_{a_1a_2...a_{n_1}},\,\,\,J_{b_{1}b_{2}...b_{n_2}} \rangle=0,\langle P_{a_1a_2...a_{n_1}},\,\,\,P_{b_{1}b_{2}...b_{n_2}} \rangle=0\nonumber \\
&&\langle P_{a_1a_2...a_{k}},\,\,\,Q_{\alpha \, a_{k+1}a_{k+2}...a_{2s+1}}\rangle=0 \nonumber \\
&&\langle Q_{\alpha \, a_1a_2...a_{k}},\,\,\,Q_{\beta \, a_{k+1}a_{k+2}...a_{2s+1}}\rangle= 4 i\epsilon_{\alpha\beta} \sum_{p}\eta_{p(a_1)p(a_2)}\eta_{p(a_3)p(a_4)}...\eta_{p(a_{2s-1})p(a_{2s})} 
\end{eqnarray}
As usual, it holds that $\langle A,\,B\rangle=(-1)^{g(A)g(B)}\langle B,\,A\rangle$ for any two generators $A$ and $B$, where
$g(A)=0$ if A is a bosonic generator and $g(A)=1$ if A is a fermionic one.
Before closing this Section, let us note that other higher-spin extensions of the Poincare group have been constructed in \cite{Savvidy:2010vb,Antoniadis:2011re}.

\subsection{Proving the Jacobi identities }
We now proceed to prove that the (anti-)commutation relations \eqref{JJcom}, \eqref{JPcom}, \eqref{PPcom},\eqref{QQcom},  \eqref{QJcom}, \eqref{QPcom} are consistent with the graded Jacobi identities
\begin{eqnarray}\label{Jacobi}
[[A,B \}, C\}+ (-1)^{g(A)(g(B)+g(C))}[[B,C\},A\}+(-1)^{g(C)(g(A)+g(B))}[[C,A\},B\}=0
\end{eqnarray}
and as a result they define a legitimate algebra.
As above, in \eqref{Jacobi} $g(A)=0$ if A is a bosonic generator and $g(A)=1$ if A is a fermionic one.

\subsubsection{Jacobi identities involving bosonic generators}
We start from the bosonic higher spin subalgebra. There are
four different types of Jacobi identities which one has to consider and which schematically take the form
$(P,P,P)$, $(P,P,J)$, $(P,J,J)$ and $(J,J,J)$. In the parenthesis one can see the generators involved.
Because the anticommutation relations of P's with themselves are zero it is obvious that the Jacobi identities  involving three P's or two P's and one J, that is ($(P,P,P)$ and  $(P,P,J)$) are trivially satisfied since each term in \eqref{Jacobi} is separately zero.

Subsequently, we consider the $(J,J,J)$ type Jacobi identity. We will show that it is satisfied by using induction. The starting point of the induction will be the case where all J's participating in the Jacobi identity  have one index. But then the Jacobi identity is certainly satisfied since all three generators belong in
the Poincare group in $2+1$ dimensions, $ISO(2,1)$.

The next step of the induction is to assume that \eqref{Jacobi} is satisfied when $A=J_{a(n_1)}$, $B=J_{b(n_2)}$ and $C=J_{c(n_3)}$ and prove that it holds when we add one additional index in one of the generators. Without loss of generality we put this additional index $a$ in $J_{a a(n_1)}$. Let us now explain our natation. By $a(n_1)$ we denote the set $a_1a_2a_3...a_{n_1}$.
Thus, $J_{a a(n_1)}=J_{a a_1a_2a_3...a_{n_1}}$. Furthermore, in what follows we will use the  notation $ \{\hat{a}_i(n_1)\}= \{a_1a_2...\hat{a}_i...a_{n_1}\}=\{a_1a_2...a_{i-1}a_{i+1}...a_{n_1}\}$. Similar notation will be used when two or more indices are missing from a set of indices, e.g.
$ \{\hat{a}_{ij}(n_1)\}= \{a_1a_2...\hat{a}_i...\hat{a}_j...a_{n_1}\}$.

We start by evaluating the first term in \eqref{Jacobi}. Before starting let us note that we will not be writing the terms where the additional index $a$ does not appear as an index in a $\epsilon$ symbol.
This is so because in all such terms the index $a$ is inert, that is, it is trivially dressing the 
relation in the previous step of the induction. As a result the sum of all such terms have to vanish.
Turning back to the first term of \eqref{Jacobi} we rewrite it as
\begin{equation}\label{Jac1}
\begin{split}
&[[J_{a a(n_1)}, J_{ b(n_2)}], J_{ c(n_3)}]=\epsilon_{a b_i}^{\,\,\,\,\,\,\,\tilde{c}}\,\,\,[J_{\tilde
{c} a(n_1)\hat{b}_i(n_2)}, J_{ c(n_3)}]+\epsilon_{a_i b_j}^{\,\,\,\,\,\,\,\tilde{c}}\,\,\,[J_{\tilde
{c} a \hat{a}_i(n_1)\hat{b}_j(n_2)}, J_{ c(n_3)}]=\\
&\epsilon_{a b_i}^{\,\,\,\,\,\,\,\tilde{c}} \epsilon_{\tilde{c}c_j d} \,\,\,J^{d}_{a(n_1)\hat{b}_i(n_2)\hat{c}_j(n_3)}+
\epsilon_{a b_i}^{\,\,\,\,\,\,\,\tilde{c}} \epsilon_{a_j c_k d} \,\,\,J^{d}_{\tilde{c} \hat{a}_j(n_1)\hat{b}_i(n_2)\hat{c}_k(n_3)}+
\epsilon_{a b_i}^{\,\,\,\,\,\,\,\tilde{c}} \epsilon_{b_j c_k d} \,\,\,J^{d}_{\tilde{c} a(n_1)\hat{b}_{ij}(n_2)\hat{c}_k(n_3)}+\\
&\epsilon_{a_i b_j}^{\,\,\,\,\,\,\,\tilde{c}} \epsilon_{a c_k d} \,\,\,J^{d}_{\tilde{c} \hat{a}_i(n_1)\hat{b}_j(n_2)\hat{c}_k(n_3)}+...,
\end{split}
\end{equation}
where as discussed above the dots denote terms in which the additional index $a$ used in the induction is not attached to an $\epsilon$ symbol.
From now on we will omit these terms altogether.
Notice that in order to keep the notation simple we have omitted the sums over $i$, $j$ and $k$.

The second term in \eqref{Jacobi} gives
 \begin{equation}\label{Jac2}
\begin{split}
[[J_{ b(n_2)},J_{ c(n_3)}], J_{a a(n_1)}]=\epsilon_{ b_i c_j}^{\,\,\,\,\,\,\,\tilde{c}}\,\,\,[J_{\tilde
{c} \hat{b}_i(n_2)\hat{c}_j(n_3)}, J_{a a(n_1)}]=
\epsilon_{ b_i c_j}^{\,\,\,\,\,\,\,\tilde{c}}\,\,\, \epsilon_{\tilde{c} a d} \,\,\,J^{d}_{a(n_1)\hat{b}_i(n_2)\hat{c}_j(n_3)}+\\
\epsilon_{ b_i c_j}^{\,\,\,\,\,\,\,\tilde{c}}\,\,\, \epsilon_{b_k a d} \,\,\,J^{d}_{\tilde{c} a(n_1)\hat{b}_{ik}(n_2)\hat{c}_j(n_3)}+
\epsilon_{ b_i c_j}^{\,\,\,\,\,\,\,\tilde{c}}\,\,\, \epsilon_{c_k a d} \,\,\,J^{d}_{\tilde{c} a(n_1)\hat{b}_{i}(n_2)\hat{c}_{jk}(n_3)}
\end{split}
\end{equation}
Finally the last term in \eqref{Jacobi} results to
\begin{equation}\label{Jac3}
\begin{split}
[[J_{ c(n_3)}, J_{a a(n_1)}], J_{ b(n_2)}]=
-\epsilon_{a c_i}^{\,\,\,\,\,\,\,\tilde{c}} \epsilon_{\tilde{c}b_j d} \,\,\,J^{d}_{a(n_1)\hat{c}_i(n_2)\hat{b}_j(n_3)}-
\epsilon_{a c_i}^{\,\,\,\,\,\,\,\tilde{c}} \epsilon_{a_j b_k d} \,\,\,J^{d}_{\tilde{c} \hat{a}_j(n_1)\hat{c}_i(n_2)\hat{b}_k(n_3)}\\-
\epsilon_{a c_i}^{\,\,\,\,\,\,\,\tilde{c}} \epsilon_{c_j b_k d} \,\,\,J^{d}_{\tilde{c} a(n_1)\hat{c}_{ij}(n_2)\hat{b}_k(n_3)}-
\epsilon_{a_i c_j}^{\,\,\,\,\,\,\,\tilde{c}} \epsilon_{a b_k d} \,\,\,J^{d}_{\tilde{c} \hat{a}_i(n_1)\hat{b}_k(n_2)\hat{c}_j(n_3)}
\end{split}
\end{equation}
Now we start assembling terms. After exchanging $i \longleftrightarrow j$ in the first term of  \eqref{Jac3} the sum of the first terms of \eqref{Jac1}, \eqref{Jac2} and \eqref{Jac3}
becomes  proportional to $\epsilon_{a b_i}^{\,\,\,\,\,\,\,\tilde{c}} \epsilon_{\tilde{c}c_j d}+\epsilon_{ b_i c_j}^{\,\,\,\,\,\,\,\tilde{c}}\,\,\, \epsilon_{\tilde{c} a d}-\epsilon_{a c_j}^{\,\,\,\,\,\,\,\tilde{c}} \epsilon_{\tilde{c}b_i d} =0$. The last equation is valid due to the fact that the usual angular momenta generators (the J's with a single index) satisfy the Jacobi identity.
The sum of the second and fourth terms in  \eqref{Jac1} and \eqref{Jac3} can be rewritten as
\begin{equation}\label{24}
\begin{split}
J_{d \tilde{c} \hat{a}_i(n_1)\hat{b}_j(n_2)\hat{c}_k(n_3)}\Big( \epsilon_{a b_j}^{\,\,\,\,\,\,\,\tilde{c}} \epsilon_{a_i c_k }^{\,\,\,\,\,\,\,\,\,d} +\epsilon_{a_i b_j}^{\,\,\,\,\,\,\,\,\,\tilde{c}} \epsilon_{a c_k }^{\,\,\,\,\,\,\,\,\,d} -
\epsilon_{a c_k}^{\,\,\,\,\,\,\,\tilde{c}} \epsilon_{a_i b_j }^{\,\,\,\,\,\,\,\,\,d} -
\epsilon_{a_i c_k}^{\,\,\,\,\,\,\,\,\,\,\tilde{c}} \epsilon_{a b_j }^{\,\,\,\,\,\,\,\,\,d}
 \Big)=0.
\end{split}
\end{equation}
In the last equation, and taking into account that J is completely symmetric under the exchange of its
indices, the first term cancels the last one while the second term cancels the third one.
Finally, the third term of \eqref{Jac1} and the second term of \eqref{Jac2} sum up to give
\begin{equation}\label{24}
\begin{split}
J_{d \tilde{c} a(n_1)\hat{b}_{ij}(n_2)\hat{c}_k(n_3)}\Big( \epsilon_{a b_i}^{\,\,\,\,\,\,\,\tilde{c}} \epsilon_{b_j c_k }^{\,\,\,\,\,\,\,\,\,d} +\epsilon_{ b_j c_k}^{\,\,\,\,\,\,\,\,\,\tilde{c}} \epsilon_{ b_i a }^{\,\,\,\,\,\,\,\,\,d}
 \Big)=0,
\end{split}
\end{equation}
while the third terms of \eqref{Jac2} and \eqref{Jac3} sum up to zero, as well
\begin{equation}\label{33}
\begin{split}
J_{d \tilde{c} a(n_1)\hat{b}_{i}(n_2)\hat{c}_{jk}(n_3)}\Big( \epsilon_{ b_i c_j}^{\,\,\,\,\,\,\,\tilde{c}} \epsilon_{ c_k a}^{\,\,\,\,\,\,\,\,\,d} -\epsilon_{ a c_k}^{\,\,\,\,\,\,\,\,\,\tilde{c}} \epsilon_{c_j b_i } ^{\,\,\,\,\,\,\,\,\,d}
 \Big)=0.
\end{split}
\end{equation}
We have, thus, proved that the Jacobi relation involving three J's is indeed, satisfied.

We now move to the case where the Jacobi identity involves 2 J's and one P generator. In this case one has to distinguish between two options. One can either put the additional index $a$ in the P generator
or in one of the two J generators. However, because the structure constants for the JJ anticommutator are precisely the same with those for the JP anticommutator the proof is identical to the one given above with the only difference that the one has to substitute P in the place of J in the right hand side of all equations from \eqref{Jac1} to \eqref{33}.

We conclude that the higher spin generators of the bosonic subalgebra with commutation relations
\eqref{JJcom},  \eqref{JPcom} and \eqref{PPcom} satisfy the Jacobi identities.

\subsubsection{Jacobi identities involving fermionic generators}
Next we move on to consider the Jacobi identities involving the fermionic generators. These fall in one of the following classes. In the first class we have the Jacobi identities with three Q's ((Q,Q,Q)-type).
These are trivially satisfied since the commutator of two supercharges Q is proportional to P which in turn anticommutes with the third supercharge to give zero for each term of \eqref{Jacobi} seperately.

In the second class, we have the Jacobi identities with two Q's and one bosonic generator which can be
either a P ((Q,Q,P)-type) or a J ((Q,Q,J)-type). As above, the (Q,Q,P)-type Jacobi identities are satisfied
because each term separately is zero. Two of the terms have a $[Q,P]$ anticommutator which is zero while the third term will be also zero since it reads $[\{Q,Q\},P]=[P,P]=0$.

In the third class belong the Jacobi identities with one supercharge and two bosonic generators. Thus one can have the following types: (Q,J,J)-type, (Q,P,J)-type and (Q,P,P)-type. The last type is trivially zero since both Q and P anticommute with P to give zero for each term of \eqref{Jacobi} separately.
The  (Q,P,J)-type is also satisfied since it gives schematically $[[Q,P],J]+[[P,J],Q]+[[J,Q],P]=0+[Q,P]+
[Q,P]=0$ since $[Q,P]=0$.
In conclusion we are left with two types of Jacobi identities that we need to evaluate one from the second class and one from the thord class, namely the (Q,Q,J)-type and the (Q,J,J)-type.

\subsubsection*{(Q,J,J)-type Jacobi Identity}

We start withe the (Q,J,J)-type Jacobi identity.
As above, we will use induction to prove  the relevant Jacobi identities.
The additional index $a$ can be added either to Q or to one of the J's.
First we consider the former case.
 \begin{equation}\label{QaJJ1}
 \begin{split}
&[[Q_{\alpha\,a a(n_1)}, J_{b(n_2)}], J_{c(n_3)}]=[J_{c(n_3)},[J_{b(n_2)},Q_{\alpha\,a a(n_1)}]]=-\frac{1}{2}
\big( \gamma_{b_i} \big)_{\alpha}^{\,\,\,\gamma}
\epsilon_{ c_j a }^{\,\,\,\,\,\,\,\,\,d}
Q_{\gamma\, d a(n_1)\hat{b}_{i} \hat{c}_{j}} +\\
&\epsilon_{ b_i a_j }^{\,\,\,\,\,\,\,\,\,d}
\epsilon_{ c_k a }^{\,\,\,\,\,\,\,\,\,e} Q_{\alpha\, d e \hat{a}_{j}(n_1)\hat{b}_{i} \hat{c}_{k}}+
 \epsilon_{ b_i a }^{\,\,\,\,\,\,\,\,\,d} \Big(\epsilon_{ c_j b_k }^{\,\,\,\,\,\,\,\,\,e} Q_{\alpha\, d e a(n_1)\hat{b}_{ik} \hat{c}_{j}}+\
 \epsilon_{ c_j a_k }^{\,\,\,\,\,\,\,\,\,e} Q_{\alpha\, d e \hat{a}_{k}(n_1)\hat{b}_{i} \hat{c}_{j}}+\\
 &\epsilon_{ c_j d }^{\,\,\,\,\,\,\,\,\,e} Q_{\alpha\,  e a(n_1)\hat{b}_{i} \hat{c}_{j}}-
 \frac{1}{2} \big( \gamma_{c_j} \big)_{\alpha}^{\,\,\,\gamma} Q_{\gamma\, d a(n_1)\hat{b}_{i} \hat{c}_{j}}
\Big)
\end{split}
\end{equation}
The second relevant term gives
\begin{equation}\label{QaJJ2}
 \begin{split}
&[[J_{b(n_2)}, J_{c(n_3)}],Q_{\alpha\, a a(n_1)} ]=
 \epsilon_{ b_i c_j }^{\,\,\,\,\,\,\,\,\,d} \Big(\epsilon_{ d a }^{\,\,\,\,\,\,\,\,\,e} Q_{\alpha\,  e a(n_1)\hat{b}_{i} \hat{c}_{j}}+
 \epsilon_{ b_k a}^{\,\,\,\,\,\,\,\,\,e} Q_{\alpha\, d e a(n_1)\hat{b}_{ik} \hat{c}_{j}}+\\
 &\epsilon_{ c_k a }^{\,\,\,\,\,\,\,\,\,e} Q_{\alpha\,  d e a(n_1)\hat{b}_{i} \hat{c}_{jk}}
 \Big)
\end{split}
\end{equation}
The third anticommutator results to
 \begin{equation}\label{QaJJ3}
 \begin{split}
&[[J_{c(n_3)},Q_{\alpha\,a a(n_1)}], J_{b(n_2)}]=\frac{1}{2}
\big( \gamma_{c_i} \big)_{\alpha}^{\,\,\,\gamma}
\epsilon_{ b_j a }^{\,\,\,\,\,\,\,\,\,d}
Q_{\gamma\, d a(n_1)\hat{b}_{j} \hat{c}_{i}} -\\
&\epsilon_{ c_i a_j }^{\,\,\,\,\,\,\,\,\,d}
\epsilon_{ b_k a }^{\,\,\,\,\,\,\,\,\,e} Q_{\alpha\, d e \hat{a}_{j}(n_1)\hat{c}_{i} \hat{b}_{k}}-
 \epsilon_{ c_i a }^{\,\,\,\,\,\,\,\,\,d} \Big(\epsilon_{ b_j c_k }^{\,\,\,\,\,\,\,\,\,e} Q_{\alpha\, d e a(n_1)\hat{c}_{ik} \hat{b}_{j}}+
 \epsilon_{ b_j a_k }^{\,\,\,\,\,\,\,\,\,e} Q_{\alpha\, d e \hat{a}_{k}(n_1)\hat{b}_{j} \hat{c}_{i}}+\\
 &\epsilon_{ b_j d }^{\,\,\,\,\,\,\,\,\,e} Q_{\alpha\,  e a(n_1)\hat{c}_{i} \hat{b}_{j}}-
 \frac{1}{2} \big( \gamma_{b_j} \big)_{\alpha}^{\,\,\,\gamma} Q_{\gamma\, d a(n_1)\hat{c}_{i} \hat{b}_{j}}
\Big).
\end{split}
\end{equation}
Now, the first term in \eqref{QaJJ1} cancels the last term of \eqref{QaJJ3},
the second term in \eqref{QaJJ1} cancels the fourth term of \eqref{QaJJ3} after changing $i \rightarrow k$, $j \rightarrow i$ and $k \rightarrow j$ in the the fourth term of \eqref{QaJJ3},
the third term in \eqref{QaJJ1} cancels the second term of \eqref{QaJJ2},
the fourth term in \eqref{QaJJ1} cancels the second term of \eqref{QaJJ3},
the last term in \eqref{QaJJ1} cancels the first term of \eqref{QaJJ3} and
the last term in \eqref{QaJJ1} cancels the third term of \eqref{QaJJ3}.
Collecting the remaining terms and reshuffling the indices $i$, $j$ and $k$ appropriately we get
\begin{equation}\label{QaJJ}
 \begin{split}
&[[Q_{\alpha\,a a(n_1)}, J_{b(n_2)}], J_{c(n_3)}]+[[J_{b(n_2)}, J_{c(n_3)}],Q_{\alpha\, a a(n_1)} ]+[[J_{c(n_3)},Q_{\alpha\,a a(n_1)}], J_{b(n_2)}]=\\
&Q_{\alpha\,  e a(n_1)\hat{b}_{i} \hat{c}_{j}}\Big(
\epsilon_{ b_i a }^{\,\,\,\,\,\,\,\,\,d}\epsilon_{ c_j d }^{\,\,\,\,\,\,\,\,\,e}+\epsilon_{ b_i c_j }^{\,\,\,\,\,\,\,\,\,d}\epsilon_{ d a}^{\,\,\,\,\,\,\,\,\,e}-\epsilon_{ c_j a }^{\,\,\,\,\,\,\,\,\,d}\epsilon_{ b_i d }^{\,\,\,\,\,\,\,\,\,e}
\Big)=0.
\end{split}
\end{equation}

Next we consider the second case, that is the $(Q,J,J)$ type Jacobi identity where the additional index, needed for the proof using the method of induction, is added to one of the generalised angular momenta generators $J$.
The first  double anticommutator we need to evaluate is
\begin{equation}\label{QJaJ1}
 \begin{split}
& [[Q_{\alpha\, a(n_1)}, J_{a b(n_2)}], J_{c(n_3)}]=-\frac{1}{2}
\big( \gamma_{a} \big)_{\alpha}^{\,\,\,\gamma}
\Big(- \frac{1}{2} \big( \gamma_{c_i} \big)_{\gamma}^{\,\,\,\delta} Q_{\delta\, d a(n_1)b(n_2)\hat{c}_{i}}  +
\epsilon_{ c_i a_j }^{\,\,\,\,\,\,\,\,\,d} Q_{\gamma\,  d \hat{a}_{j} b(n_2)\hat{c}_{i}}+\\
&\epsilon_{ c_i b_j }^{\,\,\,\,\,\,\,\,\,d} Q_{\gamma\,  a(n_1) \hat{b}_{j} \hat{c}_{i}}
\Big) +
\Big(- \frac{1}{2} \big( \gamma_{c_j} \big)_{\alpha}^{\,\,\,\delta} Q_{\delta\, d \hat{a}_{i}b(n_2)\hat{c}_{j}}  +\epsilon_{ c_j b_k }^{\,\,\,\,\,\,\,\,\,e} Q_{\alpha\,  d e \hat{a}_{i} \hat{b}_{k}\hat{c}_{j}}+
\epsilon_{ c_j a_k }^{\,\,\,\,\,\,\,\,\,e} Q_{\alpha\,  d e \hat{a}_{ik} b(n_2)\hat{c}_{j}}+\\
&\epsilon_{ c_j d }^{\,\,\,\,\,\,\,\,\,e} Q_{\alpha\,   e \hat{a}_{i} b(n_2)\hat{c}_{j}}
\Big)\epsilon_{ a a_i }^{\,\,\,\,\,\,\,\,\,d}+
\epsilon_{ c_k a }^{\,\,\,\,\,\,\,\,\,e}
\epsilon_{ b_i a_j }^{\,\,\,\,\,\,\,\,\,d} Q_{\alpha\, d e \hat{a}_{j}(n_1)\hat{c}_{k} \hat{b}_{i}}-
\frac{1}{2}
\big( \gamma_{b_i} \big)_{\alpha}^{\,\,\,\gamma}  \epsilon_{ c_j a }^{\,\,\,\,\,\,\,\,\,d}
Q_{\gamma\, d  a(n_1)\hat{c}_{j} \hat{b}_{i}}
\end{split}
\end{equation}
The second relevant double anticommutator results to
\begin{equation}\label{QJaJ2}
 \begin{split}
&[[J_{a b(n_2)}, J_{c(n_3)}],Q_{\alpha\,  a(n_1)} ]=\Big( -\frac{1}{2} \big( \gamma_{a} \big)_{\alpha}^{\,\,\,\gamma} Q_{\gamma\, d a(n_1)\hat{b}_{i}\hat{c}_{j}}
+\epsilon_{ a a_k }^{\,\,\,\,\,\,\,\,\,e} Q_{\alpha\,  d e \hat{a}_{k} \hat{b}_{i}\hat{c}_{j}}\Big)\epsilon_{ b_i c_j }^{\,\,\,\,\,\,\,\,\,d}+\\
&\epsilon_{ a c_i }^{\,\,\,\,\,\,\,\,\,d} \Big(  \epsilon_{ b_j a_k }^{\,\,\,\,\,\,\,\,\,e} Q_{\alpha\,  d e \hat{a}_{k} \hat{b}_{j}\hat{c}_{i}}-
 \frac{1}{2} \big( \gamma_{b_j} \big)_{\alpha}^{\,\,\,\gamma} Q_{\gamma\, d a(n_1)\hat{b}_{j}\hat{c}_{i}} +
 \epsilon_{ c_j a_k }^{\,\,\,\,\,\,\,\,\,e} Q_{\alpha\,  d e \hat{a}_{k} b(n_2) \hat{c}_{ij}}-
 \frac{1}{2} \big( \gamma_{c_j} \big)_{\alpha}^{\,\,\,\gamma} Q_{\gamma\, d a(n_1) b(n_2) \hat{c}_{ij}} +\\
 & \epsilon_{ d a_j }^{\,\,\,\,\,\,\,\,\,e} Q_{\alpha\,   e \hat{a}_{j} b(n_2) \hat{c}_{i}}-
 \frac{1}{2} \big( \gamma_{d} \big)_{\alpha}^{\,\,\,\gamma} Q_{\gamma\,  a(n_1) b(n_2) \hat{c}_{i}}
\Big)
\end{split}
\end{equation}
Finally, the third anticommutator reads
\begin{equation}\label{QJaJ3}
 \begin{split}
&[[J_{c(n_3)},Q_{\alpha\, a(n_1)}], J_{a b(n_2)}]=-\Big(\epsilon_{ a d }^{\,\,\,\,\,\,\,\,\,e} Q_{\alpha\,   e \hat{a}_{j} b(n_2)\hat{c}_{i}}-
 \frac{1}{2} \big( \gamma_{a} \big)_{\alpha}^{\,\,\,\gamma} Q_{\gamma\, d \hat{a}_{j}b(n_2)\hat{c}_{i}}  +\epsilon_{ a c_k }^{\,\,\,\,\,\,\,\,\,e}
Q_{\alpha\,  d e \hat{a}_{j} b(n_2)\hat{c}_{ik}}+\\
&\epsilon_{ a a_k }^{\,\,\,\,\,\,\,\,\,e} Q_{\alpha\,  d e \hat{a}_{jk} b(n_2)\hat{c}_{i}}\Big)\epsilon_{ c_i a_j }^{\,\,\,\,\,\,\,\,\,d}
+\frac{1}{2}
\big( \gamma_{c_i} \big)_{\alpha}^{\,\,\,\gamma}
\Big(- \frac{1}{2} \big( \gamma_{a} \big)_{\gamma}^{\,\,\,\delta} Q_{\delta\,  a(n_1)b(n_2)\hat{c}_{i}}  +
\epsilon_{ a a_j }^{\,\,\,\,\,\,\,\,\,d} Q_{\gamma\,  d \hat{a}_{j} b(n_2)\hat{c}_{i}}+\\
&\epsilon_{ a c_j }^{\,\,\,\,\,\,\,\,\,d} Q_{\gamma\, d a(n_1) b(n_2)\hat{c}_{ij}}
\Big)
\end{split}
\end{equation}
We are now in position to simplify the sum of the last three equations by noticing that
the second term in \eqref{QJaJ1} cancels the second term of \eqref{QJaJ3},
the third term in \eqref{QJaJ1} cancels the first term of \eqref{QJaJ2},
the fourth term in \eqref{QJaJ1} cancels the penultimate term of \eqref{QJaJ3},
the fifth term in \eqref{QJaJ1} cancels the second term of \eqref{QJaJ2}
the penultimate term in \eqref{QJaJ1} cancels the third term of \eqref{QJaJ2},
the last term in \eqref{QJaJ1} cancels the fourth term of \eqref{QJaJ2},
the last term in the second line of \eqref{QJaJ1} cancels the first term in the second line of  \eqref{QJaJ3},
the fifth term in \eqref{QJaJ2} cancels the third term of \eqref{QJaJ3} and
the third from the end term in \eqref{QJaJ2} cancels the last term of \eqref{QJaJ3}.
Collecting the remaining terms and reshuffling the indices $i$, $j$ and $k$ appropriately we get
\begin{equation}\label{QJaJ}
 \begin{split}
&[[Q_{\alpha\, a(n_1)}, J_{a b(n_2)}], J_{c(n_3)}]+[[J_{a b(n_2)}, J_{c(n_3)}],Q_{\alpha\,  a(n_1)} ]+[[J_{c(n_3)},Q_{\alpha\, a(n_1)}], J_{a b(n_2)}]=\\
&Q_{\alpha\,   b(n_2)\hat{c}_{i} \hat{a}_{j}}\Big(
-\epsilon_{ c_i a_j }^{\,\,\,\,\,\,\,\,\,d}\epsilon_{ a d }^{\,\,\,\,\,\,\,\,\,e}+\epsilon_{ a c_i }^{\,\,\,\,\,\,\,\,\,d}\epsilon_{ d a_j}^{\,\,\,\,\,\,\,\,\,e}+\epsilon_{ a a_j  }^{\,\,\,\,\,\,\,\,\,d}\epsilon_{ c_i d }^{\,\,\,\,\,\,\,\,\,e}\Big)\\
&-\epsilon_{ a c_i }^{\,\,\,\,\,\,\,\,\,d}\frac{1}{2} \big( \gamma_{d} \big)_{\alpha}^{\,\,\,\gamma} Q_{\gamma\,  a(n_1) b(n_2) \hat{c}_{i}} +
\frac{1}{4}\Big( \big( \gamma_{a} \big)_{\alpha}^{\,\,\,\gamma}\big(\gamma_{c_i} \big)_{\gamma}^{\,\,\,\delta} - \big( \gamma_{c_i} \big)_{\alpha}^{\,\,\,\gamma} \big( \gamma_{a} \big)_{\gamma}^{\,\,\,\delta} \Big)
Q_{\delta\,  a(n_1) b(n_2) \hat{c}_{i}} =0.
\end{split}
\end{equation}
In the last equation we have used the identity (see \eqref{gammadown})
\begin{equation}\label{idgammas}
 \big( \gamma_{a} \big)_{\alpha}^{\,\,\,\gamma} \big(\gamma_{c_i} \big)_{\gamma}^{\,\,\,\delta} - \big( \gamma_{c_i} \big)_{\alpha}^{\,\,\,\gamma}\big(\gamma_{a} \big)_{\gamma}^{\,\,\,\delta}
=2 \epsilon_{ a c_i }^{\,\,\,\,\,\,\,\,\,d}\big( \gamma_{d} \big)_{\alpha}^{\,\,\,\delta}
\end{equation}
to show that the last line of \eqref{QJaJ} sums to zero while the second line of the same equation is also zero due to the fact that the usual
angular momenta $J_a$ satisfy the Jacobi identity.

\subsubsection*{(Q,Q,J)-type Jacobi Identity}
The next Jacobi identity  we consider is that of two higher spin supercharges Q and one higher spin angular momentum J.
We will be using again the method of induction to prove the validity of the Jacobi identities. The additional index $a$ can be added either to one of the Q's or
to the J. First we focus on the former case.
The first double (anti-)commutator gives
 \begin{equation}\label{QaQJ1}
 \begin{split}
&[\{Q_{\alpha\, a(n_1)a }, Q_{ \beta \, b(n_2)}\}, J_{c(n_3)}]=
2 i \big( \gamma^{\tilde{c}} \big)_{\alpha\beta}[P_{\tilde{c} a(n_1) b(n_2)},J_{c(n_3)}]-
2 i \big( \gamma_{a} \big)_{\alpha\beta}[P_{ a(n_1) b(n_2)},J_{c(n_3)}]-\\
&2 i \big( \gamma_{a_i} \big)_{\alpha\beta}[P_{a \hat{a}_i(n_1) b(n_2)},J_{c(n_3)}]-
2 i \big( \gamma_{b_i} \big)_{\alpha\beta}[P_{a \hat{b}_i (n_2)a(n_1)},J_{c(n_3)}]-
4 i \epsilon_{\alpha\beta} \epsilon_{a b_i}^{\,\,\,\,\,\,\,\,\tilde{c}}[P_{\tilde{c} \hat{b}_i(n_2)a(n_1)},J_{c(n_3)}]-\\
&4 i \epsilon_{\alpha\beta} \epsilon_{a_i b_j}^{\,\,\,\,\,\,\,\,\tilde{c}}[P_{\tilde{c} a  \hat{a}_i(n_1)\hat{b}_j(n_2)},J_{c(n_3)}]=-2 i \big( \gamma^{\tilde{c}} \big)_{\alpha\beta}\epsilon_{c_i a}^{\,\,\,\,\,\,\,\,d}P_{\tilde{c} d  a(n_1)b(n_2)\hat{c}_i}-2 i \big( \gamma_{a} \big)_{\alpha\beta}\epsilon_{a_i c_j }^{\,\,\,\,\,\,\,\,d}P_{ d  \hat{a}_i b(n_2) \hat{c}_j}-\\
&2 i \big( \gamma_{a} \big)_{\alpha\beta}\epsilon_{b_i c_j }^{\,\,\,\,\,\,\,\,d}P_{ d  \hat{b}_i a(n_1) \hat{c}_j}-
2 i \big( \gamma_{a_i} \big)_{\alpha\beta}\epsilon_{a c_j }^{\,\,\,\,\,\,\,\,d}P_{ d  \hat{a}_i b(n_2) \hat{c}_j}-
2 i \big( \gamma_{b_i} \big)_{\alpha\beta}\epsilon_{a c_j }^{\,\,\,\,\,\,\,\,d}P_{ d  a(n_1)\hat{b}_i  \hat{c}_j}-\\
&4 i \epsilon_{\alpha\beta} \epsilon_{a b_i}^{\,\,\,\,\,\,\,\,\tilde{c}}\epsilon_{\tilde{c} c_j}^{\,\,\,\,\,\,\,\,d}P_{ d a(n_1) \hat{b}_i  \hat{c}_j}-
4 i \epsilon_{\alpha\beta} \epsilon_{a b_i}^{\,\,\,\,\,\,\,\,\tilde{c}}\epsilon_{a_j c_k}^{\,\,\,\,\,\,\,\,d}P_{ d \tilde{c} \hat{a}_j  \hat{b}_i  \hat{c}_k}-
4 i \epsilon_{\alpha\beta} \epsilon_{a b_i}^{\,\,\,\,\,\,\,\,\tilde{c}}\epsilon_{b_j c_k}^{\,\,\,\,\,\,\,\,d}P_{ d \tilde{c} a(n_1)  \hat{b}_{ij}  \hat{c}_k}-\\
&4 i \epsilon_{\alpha\beta} \epsilon_{a_i b_j}^{\,\,\,\,\,\,\,\,\tilde{c}}\epsilon_{a c_k}^{\,\,\,\,\,\,\,\,d}P_{ d \tilde{c} \hat{a}_i  \hat{b}_j  \hat{c}_k}
\end{split}
\end{equation}
The second double (anti-)commutator gives
\begin{equation}\label{QaQJ2}
 \begin{split}
&\{[J_{c(n_3)}, Q_{ \beta \, b(n_2)}], Q_{\alpha\, a(n_1)a }\}=\frac{1}{2}\big( \gamma_{c_i} \big)_{\beta}^{\gamma}(2i)\big( \gamma_a \big)_{ \gamma\alpha}P_{   a(n_1)  b(n_2)  \hat{c}_i}+
2 i \big( \gamma_{c_i} \big)_{\beta}^{\,\,\,\,\,\,\gamma}\epsilon_{\gamma \alpha}\epsilon_{b_j a }^{\,\,\,\,\,\,\,\,d}P_{ d a(n_1)  \hat{b}_j  \hat{c}_i}+\\
&2 i \big( \gamma_{c_i} \big)_{\beta}^{\,\,\,\,\,\,\gamma}\epsilon_{\gamma \alpha}\epsilon_{c_j a }^{\,\,\,\,\,\,\,\,d}P_{ d a(n_1)  b(n_2) \hat{c}_{ij}}-
2 i \epsilon_{c_i b_j }^{\,\,\,\,\,\,\,\,d}\big( \gamma_a \big)_{\beta\alpha}P_{ d a(n_1) \hat{b}_{j} \hat{c}_{i}}-4 i \epsilon_{\beta\alpha} \epsilon_{c_i b_j}^{\,\,\,\,\,\,\,\,d}\epsilon_{d a }^{\,\,\,\,\,\,\,\,e}P_{ e  a(n_1) \hat{b}_j  \hat{c}_i}-\\
&4 i \epsilon_{\beta\alpha} \epsilon_{c_i b_j}^{\,\,\,\,\,\,\,\,d}\epsilon_{c_k a }^{\,\,\,\,\,\,\,\,e}P_{ d e  a(n_1) \hat{b}_j  \hat{c}_{ik}}-
4 i \epsilon_{\beta\alpha} \epsilon_{c_i b_j}^{\,\,\,\,\,\,\,\,d}\epsilon_{b_k a }^{\,\,\,\,\,\,\,\,e}P_{ d e  a(n_1) \hat{b}_{jk}  \hat{c}_i}.
\end{split}
\end{equation}
Finally, the third double (anti-)commutator results to
\begin{equation}\label{QaQJ3}
 \begin{split}
&\{[J_{c(n_3)},Q_{\alpha\, a(n_1)a } ], Q_{ \beta \, b(n_2)}\}=
2 i \epsilon_{c_i a }^{\,\,\,\,\,\,\,\,d}\big( \gamma^{e} \big)_{\alpha\beta}P_{ d e a(n_1) b(n_2) \hat{c}_{i}}-
2 i \epsilon_{c_i a }^{\,\,\,\,\,\,\,\,d}\big( \gamma_d \big)_{\alpha\beta}P_{ a(n_1) b(n_2) \hat{c}_{i}}-\\
&2 i \epsilon_{c_i a }^{\,\,\,\,\,\,\,\,d}\big( \gamma_{c_j} \big)_{\alpha\beta}P_{ d  a(n_1) b(n_2) \hat{c}_{ij}}-
2 i \epsilon_{c_i a }^{\,\,\,\,\,\,\,\,d}\big( \gamma_{a_j} \big)_{\alpha\beta}P_{ d  \hat{a}_{j} b(n_2) \hat{c}_{i}}-
2 i \epsilon_{c_i a }^{\,\,\,\,\,\,\,\,d}\big( \gamma_{b_j} \big)_{\alpha\beta}P_{ d  a(n_1)\hat{b}_{j}  \hat{c}_{i}}-\\
&4 i \epsilon_{\alpha\beta} \epsilon_{c_i a}^{\,\,\,\,\,\,\,\,d}\Big(
\epsilon_{d b_j }^{\,\,\,\,\,\,\,\,e}P_{  e  a(n_1) \hat{b}_j  \hat{c}_{i}}
+\epsilon_{c_j b_k }^{\,\,\,\,\,\,\,\,e}P_{ d e  a(n_1) \hat{b}_k  \hat{c}_{ij}}+
\epsilon_{a_j b_k }^{\,\,\,\,\,\,\,\,e}P_{ d e  \hat{a}_j \hat{b}_k  \hat{c}_{i}}
\Big)-
2 i \epsilon_{c_i a_j }^{\,\,\,\,\,\,\,\,d}\big( \gamma_a \big)_{\alpha\beta}P_{ d \hat{a}_j b(n_2) \hat{c}_{i}}-\\
&4 i \epsilon_{\alpha\beta} \epsilon_{c_i a_j}^{\,\,\,\,\,\,\,\,d}\epsilon_{a b_k  }^{\,\,\,\,\,\,\,\,e}P_{ d e  \hat{a}_j \hat{b}_k  \hat{c}_{i}}+
2 i \frac{1}{2}\big( \gamma_{c_i} \big)_{\alpha}^{\gamma}\big( \gamma_{a} \big)_{\gamma\beta}P_{ a(n_1) b(n_2) \hat{c}_{i}}+
2 i \epsilon_{ a b_j}^{\,\,\,\,\,\,\,\,d}\big( \gamma_{c_i} \big)_{\alpha}^{\gamma}\epsilon_{\gamma \beta}P_{ a(n_1) \hat{b}_{j} \hat{c}_{i}}
\end{split}
\end{equation}
By cancelling the first term in \eqref{QaQJ1} by the the first term of \eqref{QaQJ3},
the second term in \eqref{QaQJ1} by the the fourth from the end term of \eqref{QaQJ3},
the third term in \eqref{QaQJ1} by the  fourth from the end term of \eqref{QaQJ2},
the fourth term in \eqref{QaQJ1} by the fourth term of \eqref{QaQJ3},
the fifth term in \eqref{QaQJ1} by the fifth term of \eqref{QaQJ3},
the seventh term in \eqref{QaQJ1} by the  third from the end term of \eqref{QaQJ3},
the penultimate term in \eqref{QaQJ1} by the last term of \eqref{QaQJ2},
the last term in \eqref{QaQJ1} by the eighth term of \eqref{QaQJ3},
the second term in \eqref{QaQJ2} by the last term of \eqref{QaQJ3},
the third term in \eqref{QaQJ2} by the third term of \eqref{QaQJ3},
the penultimate term in \eqref{QaQJ2} by the seventh term of \eqref{QaQJ3},
we, finally, obtain
\begin{equation}\label{QaQJ}
 \begin{split}
&[\{Q_{\alpha\, a(n_1)a }, Q_{ \beta \, b(n_2)}\}, J_{c(n_3)}]+
\{[J_{c(n_3)}, Q_{ \beta \, b(n_2)}], Q_{\alpha\, a(n_1)a }\}+\{[J_{c(n_3)},Q_{\alpha\, a(n_1)a } ], Q_{ \beta \, b(n_2)}\}=\\
&-4i \epsilon_{\alpha\beta}P_{d  a(n_1)\hat{b}_{i} \hat{c}_{j}}\Big(
-\epsilon_{ c_j b_i }^{\,\,\,\,\,\,\,\,\,e}\epsilon_{ e a }^{\,\,\,\,\,\,\,\,\,d}+
\epsilon_{ a b_i }^{\,\,\,\,\,\,\,\,\,e}\epsilon_{ e c_j}^{\,\,\,\,\,\,\,\,\,d}+
\epsilon_{ c_j a }^{\,\,\,\,\,\,\,\,\,e}\epsilon_{e b_i  }^{\,\,\,\,\,\,\,\,\,d}\Big)+\\
&\Big(-2i\epsilon_{  c_i a}^{\,\,\,\,\,\,\,\,\,d} \big( \gamma_{d} \big)_{\alpha\beta}-
i \big( \gamma_{a}\big)_{\alpha}^{\,\,\,\gamma} \big(\gamma_{c_i} \big)_{\gamma\beta} +
i \big( \gamma_{c_i} \big)_{\alpha}^{\,\,\,\gamma} \big( \gamma_{a} \big)_{\gamma\beta} \Big)
P_{ a(n_1) b(n_2) \hat{c}_{i}} =0.
\end{split}
\end{equation}
In the last equation we have again used the identity \eqref{idgammas} to show that the last line of \eqref{QaQJ} is zero.
The second line of the same equation is also zero due to the fact that the usual
angular momenta $J_a$ satisfy the Jacobi identity.

The last Jacobi identity to be checked involves two supercharges Q and one bosonic generator J but with the additional index $a$ needed
for the induction being put to the J, i.e. $J_{a c(n_3)}$.
Consider first the following (anti-)commutator
\begin{equation}\label{QQJa1}
 \begin{split}
&[\{Q_{\alpha\, a(n_1) }, Q_{ \beta \, b(n_2)}\}, J_{a c(n_3)}]=
2 i \big( \gamma^{\tilde{c}} \big)_{\alpha\beta}\Big(\textcolor{deepcerise}{ \epsilon_{\tilde{c} a}^{\,\,\,\,\,\,\,\,d}P_{ d  a(n_1)b(n_2)c(n_3)}}+
{\red \epsilon_{a_i a}^{\,\,\,\,\,\,\,\,d}P_{ d \tilde{c} \hat{a}_i b(n_2)c(n_3)}}+\\
&{\magenta \epsilon_{b_i a}^{\,\,\,\,\,\,\,\,d}P_{ d \tilde{c} a(n_1)\hat{b}_i c(n_3)}}
\Big)-
2 i \big( \gamma_{a_i} \big)_{\alpha\beta}\Big(
{\green \epsilon_{a_j a}^{\,\,\,\,\,\,\,\,d}P_{ d \tilde{c} \hat{a}_{ij} b(n_2)c(n_3)}}+
{\blue \epsilon_{b_j a}^{\,\,\,\,\,\,\,\,d}P_{ d  \hat{a}_i\hat{b}_j c(n_3)}}
\Big)-\\
&2 i \big( \gamma_{b_i} \big)_{\alpha\beta}\Big(
\textcolor{brown(traditional)}{\epsilon_{a_j a}^{\,\,\,\,\,\,\,\,d}P_{ d  \hat{a}_{j} \hat{b}_{i}c(n_3)}}+
\textcolor{darkviolet}{\epsilon_{b_j a}^{\,\,\,\,\,\,\,\,d}P_{ d  a(n_1)\hat{b}_{ij} c(n_3)}}
\Big)
-4i\epsilon_{\alpha\beta} \epsilon_{a_i b_j}^{\,\,\,\,\,\,\,\,d}\Big(
\epsilon_{d a}^{\,\,\,\,\,\,\,\,e}P_{e  \hat{a}_{i} \hat{b}_{j} c(n_3)}+\\
&\textcolor{amethyst}{\epsilon_{a_k a}^{\,\,\,\,\,\,\,\,\,\,e}P_{ d e  \hat{a}_{ik} \hat{b}_{j} c(n_3)}}+
\textcolor{xanadu}{\epsilon_{b_k a}^{\,\,\,\,\,\,\,\,\,\,e}P_{ d e  \hat{a}_{i} \hat{b}_{jk} c(n_3)}}
\Big)
\end{split}
\end{equation}
The second relevant (anti-)commutator is
\begin{equation}\label{QQJa2}
 \begin{split}
&\{[J_{a c(n_3)}, Q_{ \beta \, b(n_2)}],Q_{\alpha\, a(n_1) }\} =
\epsilon_{a b_i}^{\,\,\,\,\,\,\,\,\tilde{c}}\Big(
{\magenta 2 i \big( \gamma^{d} \big)_{\beta\alpha}P_{ d \tilde{c}  a(n_1) \hat{b}_{i} c(n_3)}}-
\textcolor{fuzzywuzzy}{2 i \big( \gamma_{\tilde{c}} \big)_{\beta\alpha}P_{  a(n_1) \hat{b}_{i} c(n_3)}}-\\
&\textcolor{darktan}{2 i \big( \gamma_{c_j} \big)_{\beta\alpha}P_{  \tilde{c}  a(n_1) \hat{b}_{i} \hat{c}_{j}}}-
\textcolor{darkviolet}{2 i \big( \gamma_{b_j} \big)_{\beta\alpha}P_{ \tilde{c}  a(n_1) \hat{b}_{ij} c(n_3)}}-
{\blue 2 i \big( \gamma_{a_j} \big)_{\beta\alpha}P_{ \tilde{c}  \hat{a}_{j} \hat{b}_{i} c(n_3)}}-
4 i \epsilon_{\beta\alpha}\epsilon_{\tilde{c}a_j }^{\,\,\,\,\,\,\,\,d}P_{ d  \hat{a}_{j} \hat{b}_{i} c(n_3)}-\\
&\textcolor{pink}{4 i \epsilon_{\beta\alpha}\epsilon_{c_j a_k }^{\,\,\,\,\,\,\,\,d}P_{ d \tilde{c} \hat{a}_{k} \hat{b}_{i} \hat{c}_{j}}}-
\textcolor{xanadu}{4 i \epsilon_{\beta\alpha}\epsilon_{b_j a_k }^{\,\,\,\,\,\,\,\,d}P_{ d \tilde{c} \hat{a}_{k} \hat{b}_{ij} c(n_3)}}
\Big)-
\frac{1}{2}\big( \gamma_{a} \big)_{\beta}^{\,\,\,\,\,\gamma}
\Big(\textcolor{deepcerise}{2i \big( \gamma^{d} \big)_{\gamma \alpha}P_{ d   a(n_1) b(n_2) c(n_3)}}-\\
&\textcolor{lilac}{2i \big( \gamma_{c_i} \big)_{\gamma \alpha}P_{ a(n_1) b(n_2)\hat{c}_{i} }}-
\textcolor{fuzzywuzzy}{2i \big( \gamma_{b_i} \big)_{\gamma \alpha}P_{ a(n_1) \hat{b}_{i}  c(n_3)}}-
2i \big( \gamma_{a_i} \big)_{\gamma \alpha}P_{ \hat{a}_{i} b(n_2) c(n_3)}-
\textcolor{phlox}{4 i \epsilon_{\gamma\alpha}\epsilon_{c_i a_j }^{\,\,\,\,\,\,\,\,d}P_{ d  \hat{a}_{j} b(n_2)\hat{c}_{i} }}-\\
&4 i \epsilon_{\gamma\alpha}\epsilon_{b_i a_j }^{\,\,\,\,\,\,\,\,d}P_{ d  \hat{a}_{j} \hat{b}_{i} c(n_3) }
\Big)-
2 i\epsilon_{c_i b_j }^{\,\,\,\,\,\,\,\,d}\Big(
\textcolor{yellow}{\big( \gamma_{a} \big)_{\beta\alpha}P_{ d a(n_1) \hat{b}_{j}  \hat{c}_{i}}}+
\textcolor{brightlavender}{2\epsilon_{\beta\alpha}\epsilon_{a a_k }^{\,\,\,\,\,\,\,\,e}P_{ d e \hat{a}_{k} \hat{b}_{j} \hat{c}_{i} }}
\Big)+\\
&\big( \gamma_{c_i} \big)_{\beta}^{\,\,\,\,\,\gamma}
i\Big(
\textcolor{capri}{\big( \gamma_{a} \big)_{\gamma \alpha}P_{a(n_1)  b(n_2) \hat{c}_{i}}}+
\textcolor{cadetblue}{2 \epsilon_{\gamma\alpha}\epsilon_{a a_j }^{\,\,\,\,\,\,\,\,e}P_{  e \hat{a}_{j} b(n_2) \hat{c}_{i} }}
\Big)
\end{split}
\end{equation}
The last commutator to be considered is obtained from the previous one \eqref{QQJa2} by the simultaneous exchange $\alpha \longleftrightarrow \beta$ and
$a(n_1) \longleftrightarrow b(n_2)$,  
\begin{equation}\label{QQJa3}
 \begin{split}
&\{[J_{a c(n_3)}, Q_{ \alpha \, a(n_1)}],Q_{\beta\, b(n_2) }\} =
\epsilon_{a a_i}^{\,\,\,\,\,\,\,\,\tilde{c}}\Big(
{\red 2 i \big( \gamma^{d} \big)_{\alpha\beta}P_{ d \tilde{c}  b(n_2) \hat{a}_{i} c(n_3)}}-
2 i \big( \gamma_{\tilde{c}} \big)_{\alpha\beta}P_{  b(n_2) \hat{a}_{i} c(n_3)}-\\
&\textcolor{cadetblue}{2 i \big( \gamma_{c_j} \big)_{\alpha\beta}P_{  \tilde{c}  b(n_2) \hat{a}_{i} \hat{c}_{j}}}-
{\green 2 i \big( \gamma_{a_j} \big)_{\alpha\beta}P_{ \tilde{c}  b(n_2) \hat{a}_{ij} c(n_3)}}-
\textcolor{brown(traditional)}{2 i \big( \gamma_{b_j} \big)_{\alpha\beta}P_{ \tilde{c}  \hat{b}_{j} \hat{a}_{i} c(n_3)}}-
4 i \epsilon_{\alpha\beta}\epsilon_{\tilde{c}b_j }^{\,\,\,\,\,\,\,\,d}P_{ d  \hat{b}_{j} \hat{a}_{i} c(n_3)}-\\
&\textcolor{brightlavender}{4 i \epsilon_{\alpha\beta}\epsilon_{c_j b_k }^{\,\,\,\,\,\,\,\,d}P_{ d \tilde{c} \hat{b}_{k} \hat{a}_{i} \hat{c}_{j}}}-
\textcolor{amethyst}{4 i \epsilon_{\alpha\beta}\epsilon_{a_j b_k }^{\,\,\,\,\,\,\,\,d}P_{ d \tilde{c} \hat{b}_{k} \hat{a}_{ij} c(n_3)}}
\Big)-
\frac{1}{2}\big( \gamma_{a} \big)_{\alpha}^{\,\,\,\,\,\gamma}
\Big(\textcolor{deepcerise}{2i \big( \gamma^{d} \big)_{\gamma \beta}P_{ d   a(n_1) b(n_2) c(n_3)}}-\\
&\textcolor{capri}{2i \big( \gamma_{c_i} \big)_{\gamma \beta}P_{ a(n_1) b(n_2)\hat{c}_{i} }}-
2i \big( \gamma_{a_i} \big)_{\gamma \beta}P_{ b(n_2) \hat{a}_{i}  c(n_3)}-
\textcolor{fuzzywuzzy}{2i \big( \gamma_{b_i} \big)_{\gamma \beta}P_{ \hat{b}_{i} a(n_1) c(n_3)}}-
\textcolor{yellow}{4 i \epsilon_{\gamma\beta}\epsilon_{c_i b_j }^{\,\,\,\,\,\,\,\,d}P_{ d  \hat{b}_{j} a(n_1)\hat{c}_{i} }}-\\
&4 i \epsilon_{\gamma\beta}\epsilon_{a_i b_j }^{\,\,\,\,\,\,\,\,d}P_{ d  \hat{b}_{j} \hat{a}_{i} c(n_3) }
\Big)-
2 i\epsilon_{c_i a_j }^{\,\,\,\,\,\,\,\,d}\Big(
\textcolor{phlox}{\big( \gamma_{a} \big)_{\alpha\beta}P_{ d b(n_2) \hat{a}_{j}  \hat{c}_{i}}}+
\textcolor{pink}{2\epsilon_{\alpha\beta}\epsilon_{a b_k }^{\,\,\,\,\,\,\,\,e}P_{ d e \hat{b}_{k} \hat{a}_{j} \hat{c}_{i} }}
\Big)+\\
&\big( \gamma_{c_i} \big)_{\alpha}^{\,\,\,\,\,\gamma}
i\Big(
\textcolor{lilac}{\big( \gamma_{a} \big)_{\gamma \beta}P_{b(n_2)  a(n_1) \hat{c}_{i}}}+
\textcolor{darktan}{2 \epsilon_{\gamma\beta}\epsilon_{a b_j }^{\,\,\,\,\,\,\,\,e}P_{  e \hat{b}_{j} a(n_1) \hat{c}_{i} }}
\Big)
\end{split}
\end{equation}

In the last three equations \eqref{QQJa1}, \eqref{QQJa2} and \eqref{QQJa3} terms with the same colour sum up to zero. We also notice that he sum of last term in the third line of \eqref{QQJa1}, the last term in the second line of \eqref{QQJa2} and the last term in the second line of \eqref{QQJa3} sum up to zero.
Furthermore, by using \eqref{idgammas} it is straightforward to show the second term in the fourth line of \eqref{QQJa3}, the second term in the first line of \eqref{QQJa3} and the third term in the fourth line of \eqref{QQJa2} sum to give zero.
Finally, the first term in the penultimate line of \eqref{QQJa3} cancels the first term in the penultimate line of \eqref{QQJa2}.
Thus we have proven that 
\begin{equation}\label{QQJa}
 \begin{split}
[\{Q_{\alpha\, a(n_1) }, Q_{ \beta \, b(n_2)}\}, J_{a c(n_3)}]+\{[J_{a c(n_3)}, Q_{ \alpha \, a(n_1)}],Q_{\beta\, b(n_2) }\}
-\{[ Q_{ \beta \, b(n_2)},J_{a c(n_3)}],Q_{\alpha\, a(n_1) }\}
 =0
\end{split}
\end{equation}

In conclusion, we have verified that our higher-spin super-generators satisfy the corresponding 
Jacobi identities  and as a result they close properly forming a legitimate higher-spin  superalgebra.
In the next Section, we will treat this higher-spin group as a gauge group and we will write down the action,
transformation laws and  equations of motion for the fields of the theory.

\section{Action, equations of motion and transformation laws}

Once one is equipped with the higher-spin algebra of the previous Section, one can write down an action 
that is invariant under the corresponding higher-spin group.
The idea is to treat the generalised spin connection \eqref{gaugesusy} as a gauge field taking values in $shISO(2,1)$ which is the generalisation  of $ISO(2,1)$ constructed in the previous Sections.
From this gauge field  one can construct a field strength tensor which will be interpreted as the Riemann and torsion tensors in the $2+1$-dimensional gravity. Using the language of form this reads
\begin{equation}\label{fieldstrenght}
F=R=d\Omega+ \Omega \wedge \Omega.
\end{equation}

To construct an invariant action we follow \cite{Witten}.
One considers a 4-dimensional space-time $M_4$ with a 3-dimensional boundary $\partial M_4=M_3$.
Then one can write down the following action which is invariant under the novel higher-spin group of the 
previous Sections.
\begin{equation}\label{action}
S=\int_{M_4} \tr{R \wedge R}=\int_{M_3} \tr{\Omega \wedge d \Omega+\frac{2}{3} \Omega \wedge\Omega \wedge \Omega},
\end{equation}
where the trace is defined by using the invariant bilinear form of our generalised algebra
\eqref{flatformsusy}.  In the second equality of \eqref{action} we have used the fact the the 
$R \wedge R$ in 4-dimensions is topological and can be written as a total derivative. As a result, 
one can rewrite the action as an integral defined on the 3-dimensional boundary of the 4-dimensional 
space-time. In the right hand side of \eqref{action},  one can immediately recognise the Chern-Simons 
action.  When this action is written in terms of the component fields the leading term in the expansion is 
the action describing gravity  in $2+1$ dimensions. So, our construction is the generalisation of the well-known fact that Einsteinian gravity in $2+1$ dimensions can be interpreted as a gauge theory with the gauge group being $ISO(2+1)$ \cite{Witten}. 

Subsequently, one can define  an infinitesimal gauge parameter which would be 
 \begin{eqnarray}\label{transform}
W= \sum_{n=1}^{\infty} \frac{1}{n!} \rho^{a_1a_2...a_n}P_{a_1a_2...a_n} +\sum_{n=1}^{\infty} \frac{1}{n!} \tau^{a_1a_2...a_n}J_{a_1a_2...a_n}+\sum_{n=0}^{\infty} \frac{1}{n!}
\theta^{\alpha\, a_1a_2...a_n}Q_{\alpha\, a_1a_2...a_n} ,
\end{eqnarray}
where $ \rho^{a_1a_2...a_n}$, $\tau^{a_1a_2...a_n}$ and $\theta^{\alpha\, a_1a_2...a_n}$ are infinitesimal gauge parameters.
The variation of the gauge field $\Omega_{\mu}$ under these infinitesimal gauge transformations is, as usual 
\begin{eqnarray}\label{transformomega}
\delta \Omega_{\mu}= D_{\mu} W=\partial_{\mu}W+[ \Omega_{\mu}, W],
\end{eqnarray}
where the anti-commutator is given in terms of the (anti-)commutators defining our generalised superalgebra.
In terms of componets the gauge transformations of the gauge fields read
\begin{eqnarray}\label{gauge-e}
\begin{split}
&\delta e_{\mu}^{a(n)}= \partial_{\mu} \rho^{a(n)}+\sum_{i=1}^{n}\epsilon_{b_1b_2}^{\,\,\,\,\,\,\,\,a_i}\sum_{n_1=0}^{n-1}\sum_{p}\Big(  \omega_{\mu}^{b_1 a_p(n_1)}\rho^{b_2a_p(n-n_1-1)}- 
\tau^{b_1 a_p(n_1)}e_{\mu}^{b_2a_p(n-n_1-1)}
\Big)+\\
& 2 i \sum_{i=1}^n \sum_{n_1=0}^{n-1}\sum_p \psi_{\mu}^{\gamma\, a_p(n_1)}
\big( \gamma^{a_i} \big)_{\gamma\delta} \theta^{\delta\, a_p(n-n_1-1)}
-2 i
\sum_{n_1=0}^{n}\sum_p \psi_{\mu}^{\gamma\, b a_p(n_1)}
\big( \gamma_{b} \big)_{\gamma\delta} \theta^{\delta\, a_p(n-n_1)}-\\
&2i
\sum_{n_1=0}^{n}\sum_p \psi_{\mu}^{\gamma\,  a_p(n_1)}
\big( \gamma_{b} \big)_{\gamma\delta} \theta^{\delta\,b a_p(n-n_1)}-
4i  \sum_{i=1}^{n}\epsilon_{b_1b_2}^{\,\,\,\,\,\,\,\,a_i}\sum_{n_1=0}^{n-1}\sum_{p} 
\psi_{\mu\, \delta}\,^{ b_1 \, a_p(n_1)}
 \theta^{\delta\, b_2 \,a_p(n-n_1-1)},
\end{split}
\end{eqnarray}
The generalised spin-connection transforms as 
\begin{eqnarray}\label{gauge-omega}
\begin{split}
&\delta\omega_{\mu}^{a(n)}= \partial_{\mu} \tau^{a(n)}+\sum_{i=1}^{n}\epsilon_{b_1b_2}^{\,\,\,\,\,\,\,\,a_i}\sum_{n_1=0}^{n-1}\sum_{p} \omega_{\mu}^{b_1 a_p(n_1)}\tau^{b_2a_p(n-n_1-1)}
\end{split}
\end{eqnarray}
Lastly, the generalised gravitinos transform as
\begin{eqnarray}\label{gauge-psi}
\begin{split}
&\delta \psi_{\mu}^{\alpha\,a(n)}= \partial_{\mu} \theta^{\alpha \,a(n)}+
\sum_{i=1}^{n}\epsilon_{b_1b_2}^{\,\,\,\,\,\,\,\,a_i}\sum_{n_1=0}^{n-1}\sum_{p} \Big( \omega_{\mu}^{b_1 a_p(n_1)}\theta^{\alpha\, b_2a_p(n-n_1-1)}-\tau^{b_1 a_p(n_1)}\psi_{\mu}^{\alpha\, b_2a_p(n-n_1-1)}
\Big)\\
&-\frac{1}{2}\sum_{p} \Big(\omega_{\mu}^{b_1 a_p(n_1)}\theta^{\beta\, a_p(n-n_1)}-\tau^{b_1 a_p(n_1)} \psi_{\mu}^{\beta\, a_p(n-n_1-1)}\Big)\big( \gamma_{b_1} \big)_{\beta}^{\,\,\,\,\,\,\alpha}.
\end{split}
\end{eqnarray}
A word of explanation is in order.
The sum over $p$ appearing in the first, second and last term of \eqref{gauge-e} is a sum over all inequivalent ways of splitting 
the set of  the $n-1$ indices $a_1 a_2...\hat{a}_i...a_n$ into two sets, one containing $n_1$ symmetric indices that is denoted by $a_p(n_1)$ and one containing $n-n_1-1$ symmetric indices that is denoted by $a_p(n-n_1-1)$. As usual the hat over an index is to remind us that the hatted index is absent from the set, while $a_p(n_1)$ denotes a set of $n_1$ symmetrised indices. Similarly, the sum over $p$ appearing in the third and fourth term of 
\eqref{gauge-e} is a sum over all inequivalent ways of splitting 
the full set of  the $n$ indices $a_1 a_2...a_n$ into two sets.
Similar notations have been used in \eqref{gauge-omega} and \eqref{gauge-psi}.

Under the gauge transformation \eqref{transformomega} the generalised curvature tensor $R$ transforms as
\begin{eqnarray}\label{transformR}
\delta R= [R,W].
\end{eqnarray}
Because of the cyclicity of the trace it is straightforward to show that the Chern-Simons action \eqref{action} is invariant under the gauge transformations \eqref{transformomega}, \eqref{transformR} which are based on 
the  extended super-algebra of \eqref{JJcom},  \eqref{JPcom},\eqref{PPcom}, \eqref{QQcom}, \eqref{QJcom}, \eqref{QPcom}.

We are now in position to write down the equations of motion for the fields of the theory which can be obtained from the variation of the action \eqref{action} with respect to $\Omega_{\mu}$. These are
\begin{eqnarray}\label{Reom}
R_{\mu \nu}=0 \Leftrightarrow \partial_{\mu} \Omega_\nu- \partial_{\nu} \Omega_\mu+ [\Omega_\mu, \Omega_\nu]=0
\end{eqnarray}
In term of components this equation decomposes to
\begin{eqnarray}\label{Reom1}
\begin{split}
&\partial_{\mu} e_{\nu}^{a(n)}-\partial e_{\nu}e_{\mu}^{a(n)}+ \sum_{i=1}^{n}\epsilon_{b_1b_2}^{\,\,\,\,\,\,\,\,a_i}\sum_{n_1=0}^{n-1}\sum_{p} \big(\omega_{\mu}^{b_1 a_p(n_1)}e_{\nu}^{b_2a_p(n-n_1-1)}-\omega_{\nu}^{b_1 a_p(n_1)}e_{\mu}^{b_2a_p(n-n_1-1)}\big)+ \\
&2 i \sum_{i=1}^n \sum_{n_1=0}^{n-1}\sum_p \psi_{\mu}^{\gamma\, a_p(n_1)}
\big( \gamma^{a_i} \big)_{\gamma\delta} \psi_{\nu}^{\delta\, a_p(n-n_1-1)}-2 i
\sum_{n_1=0}^{n}\sum_p \psi_{\mu}^{\gamma\, b a_p(n_1)}
\big( \gamma_{b} \big)_{\gamma\delta} \psi_{\nu}^{\delta\, a_p(n-n_1)}-\\
&2i
\sum_{n_1=0}^{n}\sum_p \psi_{\mu}^{\gamma\,  a_p(n_1)}
\big( \gamma_{b} \big)_{\gamma\delta} \psi_{\nu}^{\delta\,b a_p(n-n_1)}-
4i  \sum_{i=1}^{n}\epsilon_{b_1b_2}^{\,\,\,\,\,\,\,\,a_i}\sum_{n_1=0}^{n-1}\sum_{p} 
\psi_{\mu\, \delta}\,^{ b_1 a_p(n_1)}
 \psi_{\nu}^{\delta\, b_2 a_p(n-n_1-1)}=0,
\end{split}
\end{eqnarray}
\begin{eqnarray}\label{Reom2}
\partial_{\mu} \omega_{\nu}^{a(n)}-\partial_{\nu}\omega_{\mu}^{a(n)}+ \sum_{i=1}^{n}\epsilon_{b_1b_2}^{\,\,\,\,\,\,\,\,a_i}\sum_{n_1=0}^{n-n_1-1}\sum_{p} \omega_{\mu}^{b_1 a_p(n_1)}\omega_{\nu}^{b_2a_p(n-n_1-1)}=0
\nonumber\\
\end{eqnarray}
\begin{eqnarray}\label{Reom3}
\begin{split}
&\partial_{\mu} \psi_{\nu}^{\alpha\,a(n)}-\partial_{\nu}\psi_\mu^{\alpha\, a(n)}+ \\&\sum_{i=1}^{n}\epsilon_{b_1b_2}^{\,\,\,\,\,\,\,\,a_i}\sum_{n_1=0}^{n-n_1-1}\sum_{p} \big(\omega_{\mu}^{b_1 a_p(n_1)}\psi_{\nu}^{\alpha\, b_2a_p(n-n_1-1)}-\omega_{\nu}^{b_1 a_p(n_1)}\psi_{\mu}^{\alpha\,b_2a_p(n-n_1-1)}\big)-\\
&\frac{1}{2}\sum_{p} \big(\omega_{\mu}^{b_1 a_p(n_1)}\psi_{\nu}^{\beta\, a_p(n-n_1)}\big( \gamma_{b_1} \big)_{\beta}^{\,\,\,\,\,\alpha}-
\omega_{\nu}^{b_1 a_p(n_1)}\psi_{\mu}^{\beta\, a_p(n-n_1)}\big( \gamma_{b_1} \big)_{\beta}^{\,\,\,\,\,\alpha}
\big)=0.
\end{split}
\end{eqnarray}
As in the case of the gauge transformations of the componets of the gauge field $\Omega_{\mu}$
the sum over $p$ appearing in the first, second and last term of \eqref{Reom1} is a sum over all inequivalent ways of splitting 
the set of  the $n-1$ indices $a_1 a_2...\hat{a}_i...a_n$ into two sets, one containing $n_1$ symmetric indices that is denoted by $a_p(n_1)$ and one containing $n-n_1-1$ symmetric indices that is denoted by $a_p(n-n_1-1)$. As usual the hat over an index is to remind us that the hatted index is absent from the set, while $a_p(n_1)$ denotes a set of $n_1$ symmetrised indices. Similarly, the sum over $p$ appearing in the third and fourth term of \eqref{Reom1} is a sum over all inequivalent ways of splitting 
the set of  the $n$ indices $a_1 a_2...a_n$ into two sets.

We close this Section with two important comments.
Please notice that in the equations of motion \eqref{Reom1},  \eqref{Reom2} and \eqref{Reom3} the sum over $n_1$  has an upper bound. 
This means that only a finite number of the higher-spin fields participate in the non-linear term of the equations of motion for a given number of free indices. This is in contradistinction to the higher spin algebras shs(2) (see 
equations (38-40) of \cite{Ble}) where the  corresponding sums
run from zero to infinity and as a result the non-linear term of the equations of motion involves higher-spin fields with an unbounded number of indices. 
This difference in the "dynamics" resides in the different structure constants of the two algebras.\\
A second comment concerns the structure of \eqref{Reom1}. As it happens in ordinary gravity the generalised spin connection $\omega_{\mu}^{a(n)}$ can be considerd as an auxiliary field which can be expressed in terms of the 
vielbein $e_{\mu}^{a(n)}$ and the fermions by solving \eqref{Reom1} for $\omega_{\mu}^{a(n)}$.
Then one should substitute this solution in \eqref{Reom2} and \eqref{Reom3} to get equations for 
the vielbein and the fermions. Since the equations of motion are considerably simpler in our case 
this programme should be implemented more easily than in the cases that already exist in the literature.\\
The last comment concerns the free (linear) limit of the massless higher-spin field equations for the gauge fields. Since we are expanding around flat space the only non-zero background field is $e_{\mu}^a=\delta_{\mu}^a$. Consequently, the linearised equations of motion become
\begin{eqnarray}\label{Reomfree}
\begin{split}
&\partial_{\mu} e_{\nu}^{a(n)}-\partial e_{\nu}e_{\mu}^{a(n)}+ \sum_{i=1}^{n}\epsilon_{b_1b_2}^{\,\,\,\,\,\,\,\,a_i} \big(\omega_{\mu}^{b_1 \hat{a}_i(n-1)}e_{\nu}^{b_2}-\omega_{\nu}^{b_1 \hat{a}_i(n-1)}e_{\mu}^{b_2}\big)=0\\
&\partial_{\mu} \omega_{\nu}^{a(n)}-\partial_{\nu}\omega_{\mu}^{a(n)}=0\\
&\partial_{\mu} \psi_{\nu}^{\alpha\,a(n)}-\partial_{\nu}\psi_\mu^{\alpha\, a(n)}=0.
\end{split}
\end{eqnarray}
The hat, as usual, denotes that the corresponding index is absent from a set of indices.
It should also be noted that as in the case of pure gravity the higher-spin gauge fields do not have propagating degrees of freedom and the equations of motion \eqref{Reom} admit only trivial solutions if one restricts oneself in  a topologically trivial space. Despite this, there is a number of interesting questions in the context of the $2+1$ higher-spin theory we have constructed.
One can look for BTZ-like black hole solutions or try to introduce interactions of the higher-spin fields with matter in which case the higher-spin field may become dynamical.
Finally, one may try to extend our higher-spin algebra in $d+1$ dimensions. This will be the first step towards writing down new non-linear equations of motion and/or new invariant actions for the higher-spin gauge fields in any number of dimensions.
\section{Including a cosmological constant}
In this Section, we discuss the extension of our previous construction in the case where the background 
spacetime is $AdS_3$ or $dS_3$, that is when we have a negative ($\Lambda=-\lambda<0$) or positive ($\Lambda=-\lambda>0$) cosmological constant.
Here we will restrict ourselves to the case with no supersymmetry\footnote{Apparently, it is not straightforward to include supercharges because the supercharges will no longer anti-commute with the P-type higher-spin generators making the Jacobi identity involving three higher-spin supercharges difficult to satisfy. It would be interesting to see how one can circumvent this problem.}.
As we saw in the previous Section, higher-spin three-dimensional gravity without a cosmological constant was related to the higher-spin extension of the $ISO(2,1)$ gauge group, extension which we have constructed explictely. In a similar fashion, higher-spin gravity with a positive or negative cosmological constant $\Lambda$ should be related to some higher-spin extension of the $SO(3,1)$ or
$SO(2,2)$ groups which are the symmetry groups of the $dS_3$ and $AdS_3$ spaces respectively. We will be calling these higher-spin algebras  \eqref{JpmJpm} $hSO(3,1)$ and $hSO(2,2)$ respectively.\\
The most economical way to write down these higher-spin extensions is in terms of the following linear combinations of the angular momentum and momentum generators $J^{\pm}_a=\frac{1}{2}(J_a \pm \frac{1}{\sqrt{\lambda}}P_a )$. In terms of these generators the $dS_3$ and $AdS_3$ algebras can be written as \cite{Witten}
\begin{eqnarray}\label{alg-LA}
\begin{split}
[J^{+}_a,J^{+}_b ]=\epsilon_{abc} J^{+c},\,\,\,[J^{-}_a,J^{-}_b ]=\epsilon_{abc} J^{-c},\,\,\,[J^{+}_a,J^{-}_b ]=0.
\end{split}
\end{eqnarray}
Having the experience of the flat space higher-spin group, it is now straightforward to write down the 
anti-commutation relations for the cases at hand. These read
\begin{eqnarray}\label{JpmJpm}
&&[J^+_{a_1a_2...a_{n_1}}, J^+_{b_1b_2...b_{n_2}}]=
\sum_{i=1}^{n_1}\sum_{j=1}^{n_2}\epsilon_{a_i b_j c}J^{+c}_{\,\,\,\,\,\,\,\,\,\,a_1...\hat{a}_i...a_{n_1} b_1...
\hat{b}_j ...b_{n_2}} \nonumber\\
&&[J^-_{a_1a_2...a_{n_1}}, J^-_{b_1b_2...b_{n_2}}]=
\sum_{i=1}^{n_1}\sum_{j=1}^{n_2}\epsilon_{a_i b_j c}J^{-c}_{\,\,\,\,\,\,\,\,\,\,a_1...\hat{a}_i...a_{n_1} b_1...
\hat{b}_j ...b_{n_2}}\nonumber\\
&&[J^+_{a_1a_2...a_{n_1}}, J^-_{b_1b_2...b_{n_2}}]=0.
\end{eqnarray}
This higher-spin algebra admits the following invariant quadratic form
\begin{eqnarray}\label{flatform-LA}
&&\langle J^{+}_{a_1a_2...a_{k}},\,\,\,J^{+}_{a_{k+1}a_{k+2}...a_{2s}}\rangle=\frac{1}{2\sqrt{\lambda}}\sum_{p}\eta_{p(a_1)p(a_2)}\eta_{p(a_3)p(a_4)}...\eta_{p(a_{2s-1})p(a_{2s})} \nonumber \\
&&\langle J^{-}_{a_1a_2...a_{k}},\,\,\,J^{-}_{a_{k+1}a_{k+2}...a_{2s}}\rangle=-\frac{1}{2\sqrt{\lambda}}\sum_{p}\eta_{p(a_1)p(a_2)}\eta_{p(a_3)p(a_4)}...\eta_{p(a_{2s-1})p(a_{2s})} \nonumber \\
&&\langle J^{\pm}_{a_1a_2...a_{k}},\,\,\,J^{\pm}_{a_{k+1}a_{k+2}...a_{2s+1}}\rangle=0 \nonumber \\
&& \langle J^+_{a_1a_2...a_{k}},\,\,\,J^-_{a_{k+1}a_{k+2}...a_{n}} \rangle=0.
\end{eqnarray}
That this invariant bilinear form is consistent with the higher-spin algebra \eqref{alg-LA} can be shown in 
exactly the same way we showed the analogous statement for the flat space higher-spin algebra in the Appendix. 

In fact, when written in terms of the initial angular momenta and momenta of the $AdS_3$ or $dS_3$ space (remember that we have defined $J^{\pm}_{a(n)}=\frac{1}{2}(J_{a(n)} \pm \frac{1}{\sqrt{\lambda}}P_{a(n)} )$) \eqref{flatform-LA} is identical to the corresponding invariant bilinear for in flat space \eqref{flatform}.
It is obvious that one can define the spin connection according to 
 \begin{eqnarray}\label{gaugepm}
\Omega_\mu=\Omega^{+}_\mu+\Omega^{-}_\mu= \sum_{n=1}^{\infty} \frac{1}{n!}\omega^{+\, a(n)}_\mu J^{+}_{a(n)}+\sum_{n=1}^{\infty} \frac{1}{n!}\omega^{- \,a(n)}_\mu J^{-}_{\,a(n)}=\nonumber\\
\sum_{n=1}^{\infty} \frac{1}{n!}\omega^{ a(n)}_\mu J_{a(n)}+\sum_{n=1}^{\infty} \frac{1}{n!}e^{ a(n)}_\mu P_{a(n)},\\
\omega^{\pm a(n)}_\mu=\omega^{ a(n)}_\mu\pm \sqrt{\lambda} e^{a(n)}_\mu,J^{\pm}_{a(n)}=\frac{1}{2}(J_{a(n)} \pm \frac{1}{\sqrt{\lambda}}P_{a(n)} ) \nonumber
\end{eqnarray}
From this gauge field one can, as usual define the corresponding field strength given by
\begin{equation}\label{fieldstrenghtLA}
F=R=d\Omega+ \Omega \wedge \Omega=d\Omega^+ + \Omega^+ \wedge \Omega^+ +d\Omega^- + \Omega^- \wedge \Omega^-.
\end{equation}
Similar to the case of flat space-time one can define the following action
 \begin{equation}\label{action-LA}
S_\Lambda=\int_{M_3} \tr{\Omega \wedge d \Omega+\frac{2}{3} \Omega \wedge\Omega \wedge \Omega}
\end{equation}
where the trace is defined by using the invariant bilinear form of our generalised algebra
\eqref{flatform-LA}. 
This action is by construction invariant under the  infinitesimal gauge transformations
\begin{eqnarray}\label{transformomega-LA}
\delta \Omega_{\mu}= D_{\mu} W=\partial_{\mu}W+[ \Omega_{\mu}, W]=\partial_{\mu}W^+ +[ \Omega^+_{\mu}, W^+]+\partial_{\mu}W^- +[ \Omega^-_{\mu}, W^-],
\end{eqnarray}
with gauge parameters given by
 \begin{eqnarray}\label{transform-LA}
W= W^+ + W^-=\sum_{n=1}^{\infty} \frac{1}{n!} \tau^{+\, a_1a_2...a_n}J^{+}_{a_1a_2...a_n}+\sum_{n=1}^{\infty} \frac{1}{n!} \tau^{-\, a_1a_2...a_n}J^{-}_{a_1a_2...a_n}.
\end{eqnarray}
All anti-commutators are given in terms of the (anti-)commutators defining our generalised algebra \eqref{alg-LA}.
Similarly to pure gravity the fact that the anti-commutator of $J^+_{a(n)}$ with  $J^-_{a(n)}$ is zero and the fact that the scalar product of $J^+_{a(n)}$ with  $J^-_{a(k)}$ is also zero allows one to rewrite the 
Chern-Simons action of \eqref{action-LA} as the difference of two Chern-Simons actions, one for  
$\Omega^+$ and one for $\Omega^-$
\begin{equation}\label{action-LA1}
S_\Lambda=\int_{M_3} {\tilde Tr}\Big(\Omega^+ \wedge d \Omega^+ +\frac{2}{3} \Omega^+ \wedge\Omega^+ \wedge \Omega^+\Big)-\int_{M_3} {\tilde Tr}\Big(\Omega^- \wedge d \Omega^- +\frac{2}{3} \Omega^- \wedge\Omega^- \wedge \Omega^-\Big).
\end{equation}
The relative minus sign originates from the the relative minus sign between the first and second relations of \eqref{flatform-LA}. Furthermore, the tilde over the trace is to denote that in \eqref{action-LA1} the scalar product for both sectors is taken with a plus as in the first equation of \eqref{flatform-LA}.
Notice that when the action \eqref{action-LA1} is written in terms of the component fields the leading term in the expansion becomes the usual Einstein action in three dimensions
\begin{equation}\label{Ein-action}
S_\Lambda= \int_{M_3} d^3x \,\,\,\,\epsilon^{\mu \nu \kappa}\,\Big( e_{\mu a} \big( \partial_{\nu}\omega_{\kappa}^{\,\,\,\,a}- \partial_{\kappa}\omega_{\nu}^{\,\,\,\,a}\big)+ \epsilon_{ abc}e_{\mu}^{\,\,\,\,a}\omega_{\mu}^{\,\,\,\,b}\omega_{\kappa}^{\,\,\,\,c}+
\frac{1}{3}\lambda \epsilon_{ abc} e_{\mu}^{\,\,\,\,a}e_{\nu}^{\,\,\,\,b}  e_{\kappa}^{\,\,\,\,c}\Big).
\end{equation}

One can now proceed to derive the equations of motion for the full infinite set of component fields.
These can be derived from the variation of the action \eqref{action-LA1} with respect to the gauge fields $\Omega^+$ and $\Omega^-$ which results to 
\begin{eqnarray}\label{Reom-LA}
R^+_{\mu \nu}(\Omega^+)=0 \Leftrightarrow \partial_{\mu} \Omega^+_\nu- \partial_{\nu} \Omega^+_\mu+ [\Omega^+_\mu, \Omega^+_\nu]=0\nonumber\\
R^-_{\mu \nu}(\Omega^-)=0 \Leftrightarrow \partial_{\mu} \Omega^-_\nu- \partial_{\nu} \Omega^-_\mu+ [\Omega^-_\mu, \Omega^-_\nu]=0
\end{eqnarray}
In term of components these equations decompose to
\begin{eqnarray}\label{Reom-La2}
\partial_{\mu} \omega_{\nu}^{+\,a(n)}-\partial_{\nu}\omega_{\mu}^{+\,a(n)}+ \sum_{i=1}^{n}\epsilon_{b_1b_2}^{\,\,\,\,\,\,\,\,a_i}\sum_{n_1=0}^{n-n_1-1}\sum_{p} \omega_{\mu}^{+\,b_1 a_p(n_1)}\omega_{\nu}^{+\, b_2a_p(n-n_1-1)}=0
\end{eqnarray}
and similarly for the minus components $\omega_{\mu}^{-\,a(n)}$.
If one wishes one can employ the relations
\begin{eqnarray}\label{redef}
\omega_{\mu}^{a(n)}=\frac{1}{2}\big( \omega_{\mu}^{+\,a(n)}+\omega_{\mu}^{-\,a(n)}\big)\nonumber\\
e_{\mu}^{a(n)}=\frac{1}{2\sqrt{\lambda}}\big( \omega_{\mu}^{+\,a(n)}-\omega_{\mu}^{-\,a(n)}\big)
\end{eqnarray}
to express the equations of motion in terms of the usual higher-spin vielbein $e_{\mu}^{a(n)}$ and 
higher-spin connection $\omega_{\mu}^{a(n)}$
\begin{eqnarray}\label{Reom-La3}
\partial_{\mu} \omega_{\nu}^{\,a(n)}-\partial_{\nu}\omega_{\mu}^{\,a(n)}+ \sum_{i=1}^{n}\epsilon_{b_1b_2}^{\,\,\,\,\,\,\,\,a_i}\sum_{n_1=0}^{n-n_1-1}\sum_{p} \big(\omega_{\mu}^{\,b_1 a_p(n_1)}\omega_{\nu}^{\, b_2a_p(n-n_1-1)}
+\lambda e_{\mu}^{\,b_1 a_p(n_1)}e_{\nu}^{\, b_2a_p(n-n_1-1)}\big)=0\nonumber\\
\partial_{\mu} e_{\nu}^{\,a(n)}-\partial_{\nu} e_{\mu}^{a(n)}+ \sum_{i=1}^{n}\epsilon_{b_1b_2}^{\,\,\,\,\,\,\,\,a_i}\sum_{n_1=0}^{n-n_1-1}\sum_{p} \big(\omega_{\mu}^{\,b_1 a_p(n_1)} e_{\nu}^{\, b_2a_p(n-n_1-1)}
 -\omega_{\nu}^{\,b_1 a_p(n_1)}e_{\mu}^{\, b_2a_p(n-n_1-1)}\big)=0\nonumber\\.
\end{eqnarray}
As in the flat space construction notice the structural difference between \eqref{Reom-La3}  and the corresponding equations of motion based on the $shs(2)\oplus shs(2)$ of \cite{Ble}.
In the latter the full infinite set of higher-spin gauge fields are being entangled among themselves in the non-linear term appearing in any of the equations of motion. However, this is not the case in our construction \eqref{Reom-La3} where in any equation of motion only a finite number of fields are being entangled. As stressed in the flat space-time construction this fundamental difference originates from 
the different nature of the two higher-spin algebras, i.e. their different structure constants.

Before closing this Section, let us make an important comment. In a similar fashion with what happens with pure gravity our higher-spin extension of $SO(2,2)$ or $SO(3,1)$ gauge groups \eqref{JpmJpm}
admits beside \eqref{flatform-LA} a second invariant quadratic form which
\begin{eqnarray}\label{flatform-LA2}
&&\langle J^{+}_{a_1a_2...a_{k}},\,\,\,J^{+}_{a_{k+1}a_{k+2}...a_{2s}}\rangle=\frac{1}{2}\sum_{p}\eta_{p(a_1)p(a_2)}\eta_{p(a_3)p(a_4)}...\eta_{p(a_{2s-1})p(a_{2s})} \nonumber \\
&&\langle J^{-}_{a_1a_2...a_{k}},\,\,\,J^{-}_{a_{k+1}a_{k+2}...a_{2s}}\rangle=\frac{1}{2}\sum_{p}\eta_{p(a_1)p(a_2)}\eta_{p(a_3)p(a_4)}...\eta_{p(a_{2s-1})p(a_{2s})} \nonumber \\
&&\langle J^{\pm}_{a_1a_2...a_{k}},\,\,\,J^{\pm}_{a_{k+1}a_{k+2}...a_{2s+1}}\rangle=0 \nonumber \\
&& \langle J^+_{a_1a_2...a_{k}},\,\,\,J^-_{a_{k+1}a_{k+2}...a_{n}} \rangle=0.
\end{eqnarray}
One can of course employ this invariant form when defining the trace in the Chern-Simons action \eqref{action-LA}. Then one finds that the invariant action becomes an expression similar to \eqref{action-LA1}
with the only difference that the relative minus sign between the two terms of \eqref{action-LA1} becomes plus. The equations of motion derived from the variation of the latter action are still given by
\eqref{Reom-LA} and \eqref{Reom-La2}. So from a general point of view one could define his action
as any linear combination of \eqref{action-LA1} and its corresponding action with a plus relative sign.
Though at the classical level all these combinations are equivalent this will no longer be true at the quantum level \cite{Witten}. It would be very interesting to study the quantum behaviour of our extended higher-spin (super)gravities along the lines of \cite{Witten}.

\section{Conclusions}

In this article, we have constructed novel interacting higher-spin theories of (super)gravity in $2+1$ spacetime dimensions. Our construction is based on certain novel infinite dimensional higher-spin groups which  generalise the $ISO(2,1)$, $SO(2,2)$ and $SO(3,1)$ groups of flat space, $AdS_3$ and $dS_3$ spaces respectively. We also show that each of these extended groups admits a certain
invariant bilinear form which we specify. Being equipped with these higher-spin (super)groups we write
down, Chern-Simons like actions which are invariant under the aforementioned extended groups, the 
transformation laws of the gauge fields, and their equations of motion in all three cases. 
In the extension of the $ISO(2,1)$ group the theory describes and infinite tower of "particles" with integer and half-integer spins from 1 to infinity. In the extensions of the  $SO(2,2)$ and $SO(3,1)$ groups the theory accommodates all integer spins. One important feature of our construction is that the theories allow for an infinite number of "particles" per spin $s$.
Another related new feature of our theories is that only a finite number of fields are being entangled 
among themselves in the non-linear term appearing in any of the equations of motion.
To our knowledge, this is in contradistinction with the results in the existing literature \cite{Ble} where similar constructions based on the $shs(2) \oplus shs(2)$ higher-spin algebra result to equations of motions where the full infinite set of higher-spin gauge fields are being entangled. This fundamental difference in the "dynamics" of the two theories originates from the different nature of the two higher-spin algebras, i.e. their different structure constants.

Many interesting questions still remain open. 
The first concerns the quantisation of our theories along the line of \cite{Witten}. One might hope that this procedure will shed some light on the quantisation of gravity in general.
As mentioned above, similarly to what happens in the case of pure gravity the higher-spin gauge fields do not have propagating degrees of freedom and the equations of motion \eqref{Reom} admit only trivial solutions if one restricts oneself in a topologically trivial space. Despite this, there is a number of interesting questions in the context of the $2+1$ higher-spin theories we have constructed.
One can look for BTZ-like black hole solutions or try to introduce interactions of the higher-spin fields with matter in which case the higher-spin field may become dynamical.
One may also try to extend our higher-spin algebra in $d+1$ dimensions. This will be the first step towards writing down new non-linear equations of motion and/or new invariant actions for the higher-spin gauge fields in any number of dimensions.
Furthermore, it would be extremely interesting to interpret our constructions in the context of the $AdS/CFT$
correspondence. In particular, it would be important to identify which conformal field theory living in the 
two dimensional boundary of $AdS_3$ is the dual field theory of our higher-spin gravity living in the bulk of $AdS_3$. Once conjectured one can 
compute certain correlation functions, in a fashion similar to \cite{Giombi:2011rz,Georgiou}, at both free field theory ($\lambda \rightarrow 0$) and gravity to test the validity of the duality.
Finally, it would be important to identify the asymptotic symmetry algebra of the $2+1$-dimensional higher-spin gravity theories we have constructed in this work along the lines of \cite{Henneaux:2010xg}.

\appendix
\section{Appendix}
\subsection{Invariant quadratic form}
In this Appendix we prove that our superalgebra admits the invariant bilinear form of \eqref{flatformsusy}.
We start with the bosonic part of the algebra. It is enough to show that \eqref{flatformsusy} is consistent with the following relation which should hold for any three generators $A$, $B$ and $C$,
\begin{eqnarray}\label{bos-form}
\begin{split}
\langle A, \,\,\,[ B,C]_{\pm} \rangle+(-1)^{g(B)g(C)}\langle [ A,C]_{\pm},\,\,\,B\rangle=0.
\end{split}
\end{eqnarray}
In the last equation $g(A)$ denotes the grading of the operator $A$ which is 0 if the operator is bosonic and 1 if the operartor is fermionic. 
Actually, \eqref{bos-form} is the infinitesimal version of $\langle gAg^{-1},\,\,\, gBg^{-1}\rangle=
\langle A,\, B\rangle$, where $g$ is an arbitrary element of our supergroup.
It is easy to prove that \eqref{bos-form} is verified when all three generators are P-type generators.
This is so because any of the higher-spin P's anti-commute with any other P making each term of \eqref{bos-form} zero separately. We reach to the same conclusion when two of the three generators
are of the P-type. As above, each term of \eqref{bos-form} is separately zero either because it involves 
a $[P...,P...]=0$ anti-commutator or because it involves a $\langle P...,\, P...\rangle$  scalar product which is 
zero.
The next cases to be considered is when all three generators are of the J-type or when 2 of them are of the J-type and one is of the P-type. Actually it is not difficult to see that these two cases can be treated simultaneously since the P-P and J-J anti-commutators have identical structure constants.

Thus let us choose $A=J_{a(n_1)}$, $B=J_{b(n_2)}$ and $C=J_{c(n_3)}$. Then, in an obvious by now notation, one obtains
\begin{eqnarray}\label{bos1}
\begin{split}
&\langle A, \,\,\,[ B,C] \rangle=\epsilon_{b_i c_j}^{\,\,\,\,\,\,\,\,\,\,\,d}
\sum_{p}\eta_{p(d) p(a_1)...}\eta_{p(a_{n-1})p(a_{n})}\eta_{p(b_{1})p(b_{2})}
...\eta_{p(b_{i-2})p(b_{i-1})}\eta_{p(b_{i+1})p(b_{i+2})}...\\
&\eta_{p(b_{n_2-1})p(b_{n_2})}
\eta_{p(c_{1})p(c_2)}...\eta_{p(c_{j-2})p(c_{j-1})}\eta_{p(c_{j+1})p(c_{j+2})}...
\eta_{p(c_{n_3-1})p(c_{n_3})}=
\epsilon_{b_i c_j}^{\,\,\,\,\,\,\,\,\,\,\,d} \eta_{d{\hat b}_i {\hat c}_j,\,\,a(n_3)}\\
&=\epsilon_{b_i c_j a_k} \eta_{{\hat b}_i {\hat c}_j,\,\,\hat{a}_k},
\end{split}
\end{eqnarray}
where  \\
$\eta_{d {\hat b}_i {\hat c}_j,\,\,c(n_3)}=\sum_{p}\eta_{p(d) p(a_1)...}\eta_{p(a_{n-1})p(a_{n})}\eta_{p(b_{1})p(b_{2})}
...\eta_{p(b_{i-2})p(b_{i-1})}\eta_{p(b_{i+1})p(b_{i+2})}...\eta_{p(b_{n_2-1})p(b_{n_2})}\\
\eta_{p(c_{1})p(c_2)}...\eta_{p(c_{j-2})p(c_{j-1})}\eta_{p(c_{j+1})p(c_{j+2})}...
\eta_{p(c_{n_3-1})p(c_{n_3})}$. \\
The last equality of \eqref{bos1} is justified as follows. The index $d$ appearing in the $\epsilon$ symbol can be brought down by one of the $\eta$s.
Thus $d$ can be either one of the $b$ indices, one of the $c$ indices or one of the $a$ indices. The first two cases will give zero because they will involve $\epsilon_{b_i c_j b_k}+\epsilon_{b_k c_j b_i}=0$ or $\epsilon_{b_i c_j c_k}+\epsilon_{b_i c_k c_j}=0$ since the resulting expression should be symmetric in 
the exchange of any two $b$ indices or any two $c$ indices.
The second term in \eqref{bos-form} results to
\begin{eqnarray}\label{bos2}
\begin{split}
&\langle [ A,C], \,\,\, B \rangle=\epsilon_{a_i c_j}^{\,\,\,\,\,\,\,\,\,\,\,d}
 \eta_{d {\hat a}_i {\hat c}_j,\,\,b(n_3)}=
\epsilon_{a_i c_j b_k} \eta_{{\hat a}_i {\hat c}_j,\,\,\hat{b}_k}
\end{split}
\end{eqnarray}
By exchanging $i$ with $k$ in the last equation and by taking into account that $\eta_{{\hat b}_i {\hat c}_j,\,\,\hat{a}_k} =\eta_{{\hat a}_k {\hat c}_j,\,\,\hat{b}_i}$
it is straightforward to see the sum of the right hand sides of\eqref{bos1} and \eqref{bos2} is zero. A similar line of reasoning applies when one of the J-type 
generetors is substituted by a P-type one.
We have, thus, proven that our bosonic higher-spin subalgebra is consistent with \eqref{flatform}. 
Notice that the invariant quadratic form of our algebra is similar with the invariant quadratic form of the extension of the Poincare group
presented in \cite{Savvidy:2013gsa,Savvidy:2010vb}.

We now focus on the cases where one or more of the three generators appearing in \eqref{bos-form}
are fermionic ones.
The first case we will be considering is when all three generators are of the Q-type. Then because the commutator
of $\{Q...,Q...\}=P...$ and $\langle P...,\,\,Q... \rangle=0$ each term of \eqref{bos-form} is zero separately.
\eqref{bos-form} is also trivially satisfied when one of the  generators is fermionic and two bosonic. This happens because one ends up with scalar products of the form $\langle (bosonic)...,\,\,Q... \rangle$ which is zero
according to \eqref{flatformsusy}.
The only nontrivial situation is when we have two fermionic generators and one bosonic.
Again when the bosonic generator is a P-type one, each term of \eqref{bos-form} is satisfied  by since its left hand side will be proportional to $\langle \{Q...,Q...\}...,\,\,P... \rangle+\langle [Q...,P...],\,\,Q... \rangle=
\langle P.....,\,\,P... \rangle+\langle 0,\,\,Q... \rangle=0$ since according to \eqref{flatformsusy} $\langle P.....,\,\,P... \rangle=0$.
Finally when the bosonic generator is of the J-type one can follow similar steps  to the ones which led to 
the proof of the consistency condition in the case of three J-type generators given above.
The complete proof is cumbersome so instead of giving the complete proof we will consider two inequivalent  particular 
cases which, however capture all the necessary ingredients of the most generic case. 

The first case is for the following choice of generators $A=J_{\nu}$, $B=Q_{\alpha\, \mu_1}$ and $C=Q_{\beta\, \nu_1}$.  \eqref{bos-form} can then be rewritten as 
\begin{eqnarray}\label{proof1}
\begin{split}
&\langle \{Q_{\alpha\, \mu_1}, Q_{\beta\, \nu_1}\},\,\, J_{\nu}\rangle+
\langle Q_{\alpha\, \mu_1} ,\,\, [Q_{\beta \, \nu_1}, J_{\nu}]\rangle=2i \big( \gamma^{\rho} \big)_{\alpha \beta}\big( \eta_{\rho \mu_1}  \eta_{\nu \nu_1}+ \eta_{\rho \nu_1}  \eta_{\nu \mu_1}+\eta_{\rho \nu}  \eta_{\mu_1 \nu_1}\big)\\
&-2i \big( \gamma_{\mu_1} \big)_{\alpha \beta} \eta_{\nu \nu_1}-2i \big( \gamma_{\nu_1} \big)_{\alpha \beta} \eta_{\mu_1 \nu}
-4 i \epsilon_{\alpha\beta} \epsilon_{ \mu_1 \nu_1}^{\,\,\,\,\,\,\,\,\,\,\,\,\rho} \eta_{ \nu \rho}
-4i \epsilon_{\alpha\beta}  \epsilon_{\nu  \nu_1\mu_1}-2i \big( \gamma_{\nu} \big)_{ \beta \alpha} \eta_{\mu_1 \nu_1}=0.
\end{split}
\end{eqnarray}
In the last equation the first term cancels the fourth one, the second cancels the fifth one, the third term
cancels the last one while the penultimate is cancelled against the third from the end term.
The second inequivalent case is for the following choice of generators $C=J_{\nu}$, $B=Q_{\alpha\, \mu_1}$ and $A=Q_{\beta\, \nu_1}$.  \eqref{bos-form} can be rewritten as 
\begin{eqnarray}\label{proof2}
\begin{split}
&\langle Q_{\beta\, \nu_1},\,\, [Q_{\alpha \, \mu_1}, J_{\nu}] \rangle+
\langle [Q_{\beta \, \nu_1}, J_{\nu}] ,\,\,Q_{\alpha\, \mu_1} \rangle=
-4 i \epsilon_{\beta \alpha} \epsilon_{ \nu \mu_1 \nu_1}-
2i \big( \gamma_{\nu} \big)_{\alpha\beta}\eta_{\mu_1 \nu_1}\\
&+4i \epsilon_{\alpha\beta}  \epsilon_{\nu \nu_1 \mu_1}
+2i \big( \gamma_{\nu} \big)_{ \beta \alpha}\eta_{\mu_1 \nu_1}=0.
\end{split}
\end{eqnarray}
In \eqref{proof2} the first term cancels the third one while the second cancels the last one.

\subsection{Supersymmetry in $2+1$ dimensions}
In this Appendix we collect our notations and briefly review the supersymmetry algebra in $2+1$ dimensions since our generalised higher-spin superalgebra will have as a subalgebra the supersymmetric group in $2+1$ dimensions.
The space-time metric that we use is $\eta_{\mu \nu}=diag(-1,1,1)$.
The Dirac matrices which in 3-dimensions realise the commutation relation $\{\gamma_a, \gamma_b\}=2 \eta_{ab}$ can be chosen to be 
\begin{eqnarray}\label{gammamatrices}
\big( \gamma^0\big)^{\alpha}_{\,\,\,\,\,\beta}= -i \sigma^2,\,\,\, \big( \gamma^1\big)^{\alpha}_{\,\,\,\,\,\beta}=  \sigma^1,\,\,\,\big( \gamma^2\big)^{\alpha}_{\,\,\,\,\,\beta}=  \sigma^3,\,\,\,
\{\gamma_a, \gamma_b\}=2 \eta_{ab},
\end{eqnarray}
where $\sigma^1$, $\sigma^2$ and $\sigma^3$ are the three Pauli matrices.
We raise and lower spinor indices through the antisymmetric $\epsilon$ symbol as follows $\psi^\alpha=\epsilon^{\alpha\beta}\psi_{\beta}$ and $\psi_\alpha=\psi^{\beta}\epsilon_{\beta\alpha}$. In our conventions 
$\epsilon^{12}=\epsilon_{12}=1$ while for the antisymmetric $\epsilon^{abc}$ symbol with spacetime indices we define $\epsilon^{012}=1$.
The Dirac matrices with both spinors down are symmetric matrices since
\begin{eqnarray}\label{gammadown}
\big( \gamma^a\big)_{\alpha\beta}=\big( \gamma^a\big)^{\delta}_{\,\,\,\,\,\beta}\epsilon_{\delta\alpha},\,\,\,\big( \gamma^0\big)_{\alpha\beta}= -I=-\delta_{\alpha\beta},\,\,\, \big( \gamma^2\big)^{\alpha\beta}= - \sigma^3,\,\,\,\big( \gamma^2\big)^{\alpha\beta}=  \sigma^1.
\end{eqnarray}
This property has been used several times in the main text. Another very useful identity which has been used extensively is the following
\begin{eqnarray}\label{gammadown}
&&\big( \gamma_{a} \big)^{\alpha}_{\,\,\,\gamma} \big(\gamma_{b} \big)^{\gamma}_{\,\,\,\delta} - \big( \gamma_{b} \big)^{\alpha}_{\,\,\,\gamma}\big(\gamma_{a} \big)^{\gamma}_{\,\,\,\delta}
=-2 \epsilon_{ a b }^{\,\,\,\,\,\,\,\,\,c}\big( \gamma_{c} \big)^{\alpha}_{\,\,\,\delta}
\Leftrightarrow \nonumber \\
&&\big( \gamma_{a} \big)_{\alpha}^{\,\,\,\gamma} \big(\gamma_{b} \big)_{\gamma}^{\,\,\,\delta} - \big( \gamma_{b} \big)_{\alpha}^{\,\,\,\gamma}\big(\gamma_{a} \big)_{\gamma}^{\,\,\,\delta}
=2 \epsilon_{ a b }^{\,\,\,\,\,\,\,\,\,c}\big( \gamma_{c} \big)_{\alpha}^{\,\,\,\delta}.
\end{eqnarray}
Finally, for compwe give the supersymmetry algebra in $2+1$ dimensions 
\begin{eqnarray}\label{susy3d}
\begin{split}
&\{ Q_{\alpha},Q_{\beta} \}=2 i \gamma^c_{\alpha\beta}P_{c },\,\,\,\,\,\,
[J_{a},Q_{\alpha}]=-\frac{1}{2} \big(\gamma_{a}\big)_{\alpha}^ {\,\,\,\,\beta} Q_{\beta},\,\,\,\,\,\,
[P_{a}, Q_{\alpha}]=0 \\
&[J_a,J_b]=\epsilon_{abc} J^c,\,\,\,\,\,\,[J_a,P_b]=\epsilon_{abc} P^c,\,\,\,\,\,\,[P_a,P_b]=0.
\end{split}
\end{eqnarray}

\vspace{1cm}

\noindent {\large {\bf Acknowledgments}}

\vspace{3mm}

\noindent
We wish to thank George Savvidy  for useful discussions. 


\end{document}